\documentclass[12pt,aps,noshowpacs]{revtex4}
\usepackage{epsfig,amssymb,amsfonts,bm}

\newcommand{\vep}{{\bm p}}
\newcommand{\veq}{{\bm q}}
\newcommand{\vept}{\tilde{\bm p}}
\newcommand{\veqt}{\tilde{\bm q}}
\newcommand{\vp}{\varphi}
\newcommand{\be}{\begin{equation}}
\newcommand{\ee}{\end{equation}}
\newcommand{\ds}{\displaystyle}

\makeatletter
\renewcommand{\@makefntext}[1]{\parindent=1em\noindent\hbox to
1.8em{\hss$^{\@thefnmark}$}#1}
\renewcommand{\@footnotemark}{\hbox{\mathsurround=0pt$^{\@thefnmark}$}}
\newcommand{\ftnote}[2]{\footnotemark[#1]\footnotetext[#1]{#2}}
\makeatother
\DeclareMathSymbol{\varGamma}{\mathord}{letters}{"00}

\begin{document}

\title{\hfill{\small\rm FZJ--IKP(TH)--2009--40} \\[1.8em]
Lineshapes for composite particles with unstable constituents}

\author{C. Hanhart}
\affiliation{Forschungszentrum J\"ulich, \\ Institut f\"ur Kernphysik (IKP-3) and J\"ulich Center for Hadron
Physics, \\ Institute for Advanced Simulation (IAS-4),\\
D-52425 J\"ulich, Germany}

\author{Yu. S. Kalashnikova and A. V. Nefediev}
\affiliation{Institute for Theoretical and Experimental Physics, 117218,
B.Cheremushkinskaya 25, Moscow, Russia}

\begin{abstract}
We discuss on very general grounds possible lineshapes of composite
particles with one unstable constituent. Expressions are
derived in a coupled-channel formalism for constituents
interacting in an $S$-wave with no assumption made on the
width of one of them. We show how easy-to-use formulae, already existing in the
literature, emerge for narrow constituents and identify the
parameter that controls the regime of their applicability.
\end{abstract}

\maketitle

\section{Introduction}

In recent years charm spectroscopy had a renaissance as a result of a huge
amount of data that came from modern experiments.
Many new states were found at $B$-factories and Tevatron, with the bulk of
them showing properties in  vast conflict
with what was expected from quark models
of the Cornell type \cite{cornell} or variants
thereof --- for a recent review see Ref.~\cite{olsen}. The experimental 
situation is expected to evolve further with BES-III just
having started  data taking and with  LHCb and PANDA on the horizon.

If there are no nearby thresholds, the lineshapes of resonances may be
presented in the standard Breit--Wigner form. However, the proximity
of a (relevant) threshold distorts the lineshape strongly, as was
stressed recently in studies of the $X(3872)$, and then the lineshape of a
resonance may contain important information on its nature
\cite{braaten1,recon} --- under certain conditions, discussed
below, this enables one to distinguish between genuine
quarkonium states and states with a significant admixture of a
hadronic molecule in the wave function.  However, it was found in
Refs.~\cite{woolly,braaten1,braaten2} that the resonance lineshape
might well get distorted significantly by the width of the
constituents. 
In this paper we investigate on very general grounds how this
distortion depends on the width of the constituents and under
which circumstances simplified formulae for the line shapes can be
used.

Indeed, the list of new near-threshold states includes, together with
the above-mentioned $X(3872)$ particle, located close to the
$\bar DD^*$ threshold, such entries as the $Y(4260)$
\cite{4260} (which is not far from the $D \bar D_1$ threshold, where
$D_1$ is a $P$-wave $1^{+}$ charmed meson) and the $Y(4660)$ state \cite{Y4660exp} (close to the
$\psi' f_0(980)$ threshold). While the width of the $D^*$
is tiny (less than $100$ keV), constituent widths, generally, can
be as large as  $50$ MeV for $f_0(980)$ or even larger.
In this regard, we address the following questions:
\begin{itemize}
\item What are the most general expressions for the lineshapes with no 
assumption made on the width of the constituents?
\item What is the parameter that controls the deviations from the simplified
expressions of Refs.~\cite{woolly,braaten1}?
\end{itemize}

The formalism employed is a straightforward generalisation of 
that used in Ref.~\cite{evidence}, where only
the case of stable constituents was discussed.
We believe that our insights will prove useful for the
analysis of future experiments.

\section{Collection of Main Results}

In order to simplify the reading of the paper and to allow those interested in the
final result only to skip all technical details, we collect in this chapter
the main results and conclusions --- the derivation of those is then presented
in the subsequent chapters.  Indeed, while the derivation of the main results
turns out to be rather technical, the final conclusions are not.

\begin{center}
\begin{figure}[t]
\epsfig{file=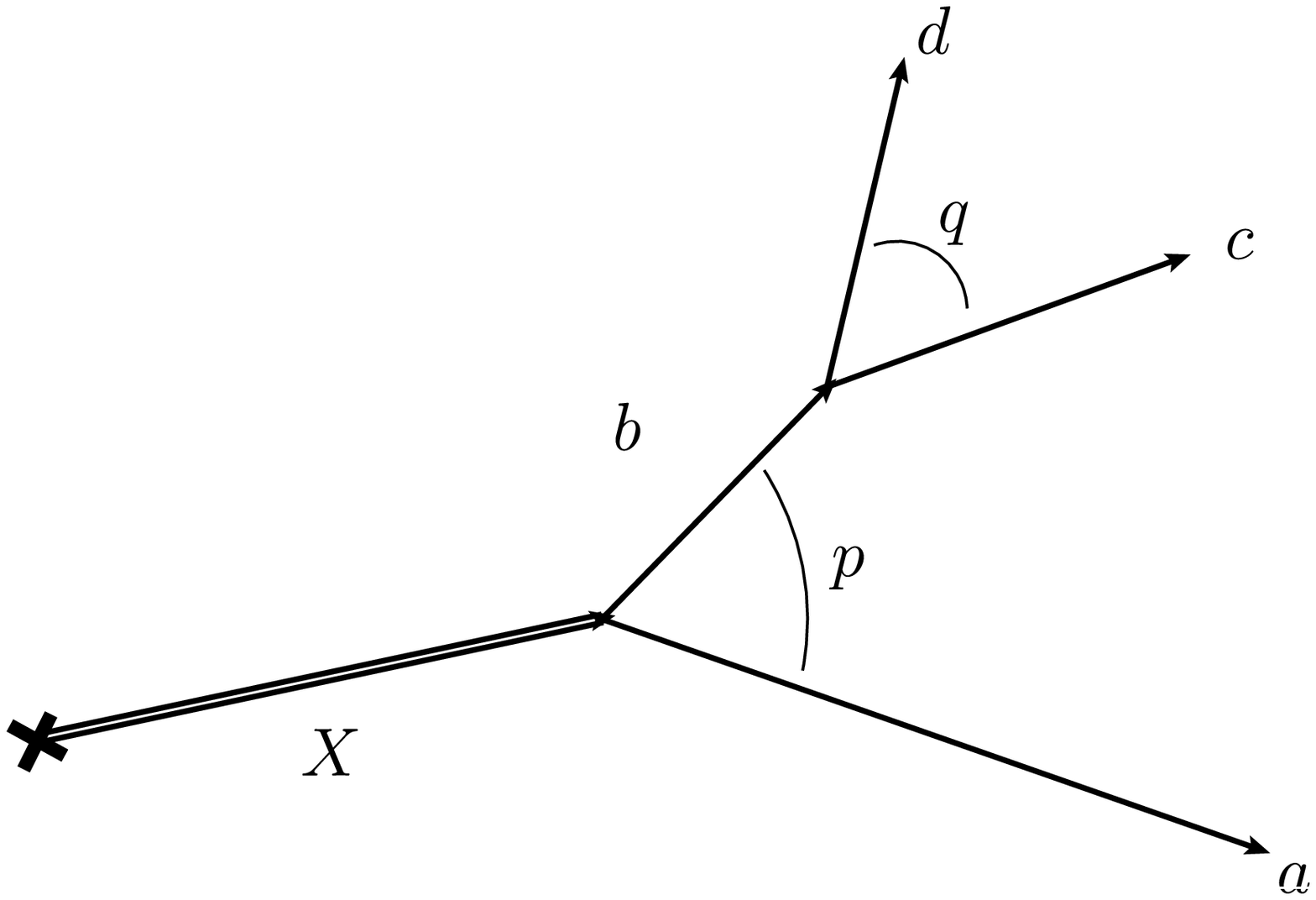, width=10cm}
\caption{Kinematics of the $X$ decay. The $X$ is assumed to be produced from a local source (denoted as a cross), and
all details of its production can be absorbed into a constant ${\cal B}$.}\label{diag}
\end{figure}
\end{center}

For simplicity we denote the decaying particle as $X$. Although the
results derived are more general and apply not only to the $X(3872)$
charmonium, we shall use the latter as a paradigmatic example
for the application of the suggested approach. We use the following
shorthand notations: in the intermediate state there are particles $a$
(stable) and $b$ (unstable, with the width $\varGamma_R$); in the final state we have the same
particle $a$, while the particle $b$ decays into particles $c$ and
$d$. Thus, in these notations, the process under consideration is
$X\to ab\to a[cd]$ (the corresponding chain for the $X(3872)$ obviously looks like 
$X(3872)\to \bar{D}D^*\to \bar{D}[D\pi]$). The masses of the particles
are labelled accordingly (see Fig.~\ref{diag}). Then, as outlined below,
in most relevant cases the differential rate close to the
$ab$ threshold, with a relative $S$-wave between $a$ and $b$, may be written as
[see Esq.~(\ref{elastic}) below]
\be 
\frac{d{\rm Br}(a[cd])}{dE}={\cal B}\frac1{2\pi}
\frac{g_{ab}k_{\rm eff}(E)}{\left|E-E_X+\frac12 g_{ab}\left[\kappa_{\rm eff}(E)
+ik_{\rm eff}(E)\right]+\frac{i}{2}\varGamma_0\right|^2},
\label{rate_final}
\ee 
where the constant ${\cal B}$ absorbs all details of the $X$ production, $g_{ab}$ denotes the effective coupling
constant of the $X$ particle to $ab$, and $\kappa_{\rm eff}(E)$ and $k_{\rm eff}(E)$ parameterise the real
and the imaginary part of the $X$ self-energy. The energy $E$ is counted from
the three-body threshold, so that the full mass is
$M=m_a+m_c+m_d+E$. Similarly, the parameter $E_X$ is related to the physical mass
of the $X$ particle as $M_X=m_a+m_c+m_d+E_X$.  In addition
Eq.~(\ref{rate_final}) contains $\varGamma_0$ to mimic the presence of extra,
inelastic channels (such as $\pi^+\pi^- J/\psi$, $\pi^+\pi^-\pi^0 J/\psi$,
{\em etc.} for the $X(3872)$). The corresponding inelastic rate then reads
[see Eq.~(\ref{inelastic}) below]
\be
\frac{d{\rm Br}(\mbox{inel})}{dE}={\cal B}
\frac1{2\pi} \frac{\varGamma_0}{\left|E-E_X+\frac12 g_{ab}
\left[\kappa_{\rm eff}(E) +ik_{\rm eff}(E)\right)]+\frac{i}{2}\varGamma_0\right|^2}.
\label{rate_final_inel}
\ee
For a vanishing width of the constituent, $\varGamma_R=0$,
Eq.~(\ref{rate_final}) reduces to the well-known
Flatt{\'e}-type expression with
\begin{eqnarray}
k_{\rm eff}(E)&=&\sqrt{2\mu_p (E-E_R)}\;\Theta(E-E_R),\label{ksim} \\
\kappa_{\rm eff}(E)&=&-\sqrt{2\mu_p (E_R-E)}\;\Theta(E_R-E)+\sqrt{2\mu_p (E_R-E_X)}\;\Theta(E_R-E_X)
\label{effmomgam0}
\end{eqnarray}
for the relative momentum of the $ab$ system and its analytic continuation
below threshold, respectively. Here $E_R$ denotes
the energy of the unstable constituent, also measured relative
to the three-body threshold, via $m_b=m_c+m_d+E_R$, and
$\mu_p$ denotes the reduced mass of the $ab$ system.
Note that, deviating from the standard expression, 
$\kappa_{\rm eff}$ was subtracted at $E=E_X$
in order to comply with the definition of $E_X$ as a
zero of the real part of the inverse scattering matrix.
One should keep in mind that all energies are defined
with respect to the three-body threshold. However, in the limit
$\varGamma_R=0$ this threshold becomes irrelevant
and the only relevant threshold is the $ab$ two-body threshold located
at $E=E_R$ --- see the discussion below Eq.~(\ref{GX}).

Introducing a nonvanishing width
$\varGamma_R$ implies a certain smearing of both $\kappa_{\rm eff}(E)$ and
$k_{\rm eff}(E)$, so that, in the presence of a finite width $\varGamma_R\neq
0$, more complicated expressions for $\kappa_{\rm eff}(E)$ and $k_{\rm eff}(E)$
need to be used, and we derive such expressions below [as given in
Eqs.~(\ref{keff})-(\ref{kappa2full})].
One might expect that for an isolated, narrow constituent, distant from any
threshold, the situation simplifies, and this is indeed what we observe.
We therefore define the dimensionless ratio
\be
\lambda=\frac{\varGamma_R}{2E_R},
\label{lambdadef}
\ee
with $E_R$ being the energy of the unstable constituent measured with respect
to the three-body threshold.  In this work we only study a resonance coupling
to one channel --- thus the constituent is isolated by construction. The
parameter $\lambda$ then captures the other two conditions, since for
$\lambda\ll 1$ the width of the constituent is much smaller than the distance
to the nearest threshold.  It will be demonstrated below that for this
situation much simpler formulae emerge --- our exact expressions reduce to
those presented in Ref.~\cite{woolly} long time ago and, more recently, in
Ref.~\cite{braaten1}, namely
\begin{eqnarray}
k^{(0)}_{\rm eff}(E)&=&\sqrt{\mu_p}
\sqrt{\sqrt{(E-E_R)^2+\varGamma^2/4}+E-E_R},
\label{kwoolly}\\ \nonumber
\kappa^{(0)}_{\rm eff}(E)&=&
-\sqrt{\mu_p}\sqrt{\sqrt{(E-E_R)^2+\varGamma^2/4}-E+E_R} \\
& & \qquad +\sqrt{\mu_p}\sqrt{\sqrt{(E_X-E_R)^2+\varGamma^2/4}-E_X+E_R},
\label{kappawoolly}
\end{eqnarray}
where, analogously to above, a subtraction was introduced for $\kappa^{(0)}_{\rm eff}$.
In Ref.~\cite{woolly} the width was taken as constant, $\varGamma\equiv\varGamma_R$.
On the other hand,
in Ref.~\cite{braaten1} the same expressions were given, but  with a 
``running" width, 
\be
\varGamma(E)=\varGamma_R\left(\frac{E}{E_R}\right)^{l+1/2},
\label{runwid}
\ee 
where $l$ is the orbital angular momentum in the subsystem $cd$.  We shall
demonstrate below that for both $S$- and $P$-wave\ftnote{1}{Notice that the
$P$-wave case is relevant for the $X(3872)$, assumed to be a $D^*\bar{D}$
bound system, with the $D^*$ decaying into $\pi D$ in a $P$-wave.}
resonances as constituent, in the case of $\lambda\ll 1$, Eqs.~(\ref{kwoolly})
and (\ref{kappawoolly}), when used in Eq.~(\ref{rate_final}), describe the
lineshapes quite accurately.  In this limit the width of the resonance can
safely be considered constant and thus there is no improvement to use an
energy-dependent width from Ref.~\cite{braaten1} instead. However, we shall
also show that already for values as small as $\lambda\approx 0.2$ significant
deviations become observable between the full expressions and the approximate
ones. Especially, for the approximate expressions, and this holds for both
Eqs.~(\ref{kwoolly}) and (\ref{kappawoolly}) with a constant as well
as with an energy-dependent width, the amplitude shows a wrong
behaviour near the three-body threshold, which is part of the signal region
for values of $\lambda\approx 0.1$ and larger.  The results for the lineshapes
are illustrated in Figs.~\ref{BrS1}-\ref{BrP2}.

We also discuss the effect of a possible interference in the final state. For example,
since the $X(3872)$ decays in equal parts via
$\bar{D}^0D^{0*}\to\bar{D}^0[D^0\pi^0]$ or $D^0\bar{D}^{0*}\to
D^0[\bar{D}^0\pi^0]$, the final states of both decay chains interfere.
We argue that, to avoid a violation of unitarity, one is to include
interference effects from the very beginning, by solving the full
coupled-channel problem. However, for the constituent width small enough, a
simple modification of Eq.~(\ref{rate_final}) is possible which accounts for
the interference effects, namely
\be
\frac{d{\rm Br}(a[cd])}{dE}={\cal B}\frac{1}{2\pi}
\frac{g_{ab}(k_{\rm eff}(E)+k_{\rm int}(E))}{(E-E_X+\frac{g_{ab}}{2}
\kappa_{\rm eff}(E))^2+\frac{1}{4}\left(\varGamma_0+g_{ab}
(k_{\rm eff}(E)+k_{\rm int}(E))\right)^2}.
\ee
For a motivation for this expression see
text near Eq.~(\ref{DDpiimproved}). Notice that the
interference contribution $k_{\rm int}(E)$, defined in
Eq.~(\ref{kintdef}), is taken into account both in the
numerator and in the denominator of the differential rate
as demanded by unitarity. We call this
``improved'' interference --- to be confronted with the ``naive'' interference
discussed in the literature, with the $k_{\rm int}(E)$ term included in the
numerator only [compare Eqs.~(\ref{DDpiimproved}) and (\ref{DDpi}) and see
Fig.~\ref{BrS1v2}].

This finalises our collection of main results. Before discussing them in
further detail we now first turn to their derivation.

\section{Generalities}

We now start with the detailed derivation of the results presented above.
For simplicity we assume that the dynamics in all subsystems can be treated nonrelativistically.
Our Fock space consists of three states: a compact seed, labelled as
$|X_0\rangle$, the $ab$ pair, and the $a[cd]$ state. Thus the wave function of
the $X$ can be written as
\be
|\Psi\rangle=
\left(
\begin{array}{c}
C|X_0\rangle\\
\chi(\vep)|ab\rangle\\
\vp(\vep,\veq)|a[cd]\rangle
\end{array}
\right).
\label{psi}
\ee

Here $\vep$ and $\veq$ are the
centre-of-mass momenta in the $ab$ and $cd$ subsystems, respectively --- see Fig.~\ref{diag}.

Wave function (\ref{psi}) obeys a Schr{\"o}dinger-like equation:
\be
H|\Psi\rangle=M|\Psi\rangle,
\label{schequation}
\ee
with the Hamiltonian
\be
H=H_0+V.
\ee

If we label the channels as
\be
1=|X_0\rangle,\quad 2=|ab\rangle,\quad 3=|a[cd]\rangle,
\ee
we may write
\be
H_{01}=M^{(0)},\quad H_{02}=m_a+m_b^{(0)}+\frac{p^2}{2\mu_{ab}},\quad
H_{03}=m_a+m_c+m_d+\frac{p^2}{2\mu_p}+\frac{q^2}{2\mu_q},
\label{H0}
\ee
with the reduced masses:
\be
\mu_{ab}=\frac{m_am_b^{(0)}}{m_a+m_b^{(0)}},
\quad\mu_p=\frac{m_a(m_c+m_d)}{m_a+m_c+m_d},
\quad\mu_q=\frac{m_cm_d}{m_c+m_d}.
\label{redm}
\ee
The masses $M^{(0)}$ and $m_b^{(0)}$ entering expressions (\ref{H0}) and (\ref{redm}) are bare masses. 

Transitions allowed are
$|X_0\rangle\leftrightarrow |ab\rangle$, mediated by the transition potential
$V_{12}=V_{21}=f(\vep)$, and
$|ab\rangle\leftrightarrow |a[cd]\rangle$ (particle $a$ being a spectator), mediated by the
transition potential $V_{23}=V_{32}=g(\veq)$.
Notice that, in addition to the transition potentials $g(\vep)$ and $f(\vep)$,
one might allow for a diagonal interaction in the $ab$ channel.
It was shown in Refs.~\cite{weinbergevidence,interplay}, however,
that nonperturbative diagonal
interactions in the mesonic channels can
in general be removed within effective theories
employing a unitary transformation --- this is not
possible only in the unusual situation of a
hadronic molecule and a compact state, both weakly coupled,
being present near threshold simultaneously \cite{interplay}. In this
way all nonperturbative meson--meson interactions
get cast into the effective coupling constants
and the parameter $M^{(0)}$. As we shall see, the latter quantity
does not play a role in the whole analysis,
while the former contain important physics information --- see Sec.~\ref{motwf}.

We shall focus on resonances near threshold. This allows us to use
nonrelativistic kinematics and to identify $\mu_{ab}=\mu_p$. The
latter implies that, in the expression for the reduced masses 
the bare mass $m_b^{(0)}$ is to be replaced by the physical
mass $m_b$ (to be defined below). 

Thus
\be
H=H_0+V=
\left(
\begin{array}{ccc}
M^{(0)}&f(\vep)&0\\
f(\vep)&m_a+m_b^{(0)}+T_{ab}(p)&g(\veq)\\
0&g(\veq)&m_a+m_c+m_d+T_{ab}(p)+T_{cd}(q)
\end{array}
\right),
\label{H}
\ee
where two convenient abbreviations were introduced:
$$
T_{ab}(p)=\frac{p^2}{2\mu_p},\quad T_{cd}(q)=\frac{q^2}{2\mu_q},
$$
for the kinetic energies of the $ab$ and the $cd$ subsystem, respectively.

Now one can define the fully off-shell $t$-matrix as a solution of the matrix
Lippmann--Schwinger equation, written symbolically as
\be
t_{ik}=V_{ik}-V_{ij}{\cal G}^{(0)}_{jl}t_{lk},
\label{equation}
\ee
where the subscript labels channels and summation over repeated indices is understood.
The free Green's function is 
\be
{\cal G}^{(0)}=(H_0-M)^{-1},\quad 
{\cal G}_{ik}^{(0)}=\delta_{ik}{\cal G}_i^{(0)}.
\ee

For convenience we express all energies with respect to the thresholds, thus
writing
\be
M=m_a+m_c+m_d+E,\quad M^{(0)}=m_a+m_c+m_d+E_0,\quad m_b^{(0)}=m_c+m_d+E_b^{(0)},
\label{E}
\ee
for the total energy $E$, the bare energy of the $X$, $E_0$, and the bare energy $E_b^{(0)}$ of the particle $b$,
respectively.

The solution of Eq.~(\ref{equation}) takes the form
\begin{eqnarray}
t_{11}(E)&=&\frac{G_X^{(0)}(E)(E-E_0)}{E-E_0+G_X^{(0)}(E)},\\
t_{21}(\vep,E)&=&t_{12}(\vep,E)=\frac{f(\vep)D_2(p,E)}{\Delta(p,E)}
\frac{E_0-E}{E-E_0+G_X^{(0)}(E)},\\
t_{31}(\vep,\veq,E)&=&t_{13}(\vep,\veq,E)=
-\frac{g(\veq)f(\vep)}{\Delta(p,E)}\frac{E_0-E}{E-E_0+G_X^{(0)}(E)},\\
t_{22}(\vep,\vep',E)&=&
\delta(\vep-\vep')\frac{G_b^{(0)}(p,E)D_2(p,E)}{\Delta(p,E)}\nonumber\\
&+&\frac{D_2(p,E)}{\Delta(p,E)}\frac{f(\vep)f(\vep')}{E-E_0+G_X^{(0)}(E)}
\frac{D_2(p',E)}{\Delta(p',E)},\\
t_{23}(\vep,\vep',\veq',E)&=&t_{32}(\vep',\veq',\vep,E)
=-\delta(\vep-\vep')\frac{g(\veq')D_2(p',E)}{\Delta(p,E)}\nonumber\\
&-&\frac{g(\veq')f(\vep)f(\vep')D_2(p,E)}{\Delta(p,E)\Delta(p',E)}
\frac{1}{E-E_0+G_X^{(0)}(E)},\\
t_{33}(\vep,\veq,\vep',\veq',E)&=&\delta(\vep-\vep')
\frac{g(\veq)g(\veq')}{\Delta(p,E)}+
\frac{g(\veq)g(\veq')f(\vep)f(\vep')}{\Delta(p,E)\Delta(p',E)}
\frac{1}{E-E_0+G_X^{(0)}(E)},
\end{eqnarray}
where the bare self-energy parts
\be
G_b^{(0)}(p,E)=\int d^3q \frac{g^2(\veq)}{T_{ab}(p)+T_{cd}(q)-E-i0},
\label{G20}
\ee
\be
G^{(0)}_X(E)=\int d^3p\frac{f^2(\vep)}{E_b^{(0)}+T_{ab}(p)-E-G_b^{(0)}(p,E)-i0},
\label{GX0}
\ee
drive the propagation of particle $b$ and the $X$, respectively, and
\be
D_2(p,E)=T_{ab}(p)+E_b^{(0)}-E,
\ee
\be
\Delta(p,E)=G_b^{(0)}(p,E)-D_2(p,E).
\ee

Let us now introduce physical masses $m_b$ and $M_X$ and the corresponding energies $E_R$ and $E_X$
as
\be
m_b=m_c+m_d+E_R,\quad M_X=m_a+m_c+m_d+E_X.
\ee
We focus on the case of particle $b$ being a resonance, thus having $E_R>0$. 

The renormalised self-energies are then written as, employing subtractions at the
physical masses (or energies $E_R$ and $E_X$),
\begin{eqnarray}
G_b(p,E)&=&G_b^{(0)}(p,E)-{\rm Re}\left(G_b^{(0)}\left(0,E_R\right)\right),\label{G2}\\
G_X(E)&=&\int d^3p\frac{f^2(\vep)}{E_R+T_{ab}(p)-E-G_b(p,E)-i0}\nonumber\\
&-&{\rm Re}\left(\int d^3p\frac{f^2(\vep)}{E_R+T_{ab}(p)-E_X-G_b(p,E_X)-i0}\right).
\label{GX}
\end{eqnarray}
In such a way,
the physical energies $E_R$ and $E_X$ are defined as zeros of the real parts of the corresponding dressed propagators.

Note that the self-energy of the $X$ has a very peculiar analytic structure~(see
Ref.~\cite{mitmichael} and references therein): since $G_b$ has a branch
point singularity at the opening of the $cd$ threshold, the $X$ self-energy
shows two branch points inside the complex plane on the unphysical sheet --- for
the unstable resonance the physical sheet is the one where all three particles
$a$, $c$, and $d$ are
physical; the two sheets meet at the three-particle threshold.
To understand the role of these singularities better, it is instructive
to observe the movement of those branch points as the width of the resonance
decreases: then they approach the physical axis while at the same time the
coupling to the physical sheet gets weaker and weaker. In the limit of 
a vanishing width of the resonance, what used to be the physical sheet decouples
completely converting what was the unphysical sheet to the physical sheet;
at the same time the branch points merge on the physical axis to produce
the branch point that starts the unitarity cut for the $ab$ channel.

Equation (\ref{schequation}) for the wave function possesses three types of
solutions (which can be defined by the corresponding asymptotic
behaviour): 
(i) bound state of all three outgoing particles ($a$, $c$, and $d$), (ii) bound 
state in the subsystem $cd$ and a free motion between this bound state and the 
particle $a$, and (iii) a free motion of all three particles, $a$, $c$, and $d$. Below we
assume that neither a three-body bound state nor a bound state in the $cd$ 
subsystem exists, and concentrate on the solution (iii), which is
the situation most relevant for the $X(3872)$ being located
close to the $D^*D$ threshold with the $D^*$ as a resonance
in the $D\pi$ channel. 

All relevant matrix elements can be found from the multichannel $t$-matrix discussed
above. However, there are no prospects for experimental studies of, say,
$\bar{D}D^*$ elastic scattering, or of the inelastic process
$\bar{D}D^*\to\bar{D}D\pi$. All information on the resonance
$X$ is therefore extracted from its production from an external source.
For example, from $B\to KD\bar{D}\pi$ in case of the $X(3872)$ particle.
Then, if the said source can be considered as pointlike, the
production amplitude may be written as\ftnote{2}{More correctly this
is the expression for the production through the quark--antiquark
component of the $X$ wave function (see the discussion in
Ref.~\cite{xnew}) or if there is only a hadronic molecule present. 
It might be distorted significantly, if a weakly
coupled molecular state and a quark state are present simultaneously
in the near-threshold region --- see Ref.~\cite{interplay}. However,
since this is a quite exotic scenario we do not consider this case
further here.}
\be 
{\cal M}=\mbox{const}\times {\cal G}_1^{(0)}(E)t_{13}(\vept,\veqt,E)=
\mbox{const}\times \frac{f(\vept)g(\veqt)}{[E-E_X+G_X(E)][E-E_R-T_{ab}(\tilde{p})+G_b(\tilde{p},E)]},
\label{prr}
\ee 
where the details of the production dynamics are absorbed into the
overall constant multiplier and the momenta $\vept$ and $\veqt$
satisfy the total energy conservation condition,
\be
E-T_{ab}(\tilde{p})-T_{cd}(\tilde{q})=0.
\ee

For future convenience it is instructive to rederive Eq.~(\ref{prr}) directly from the
solution of Schr{\"o}dinger equation (\ref{schequation}). Indeed, for the case of a free motion of all three
particles in the final state we may write
\be
\Psi_i=\Psi_i^{(0)}-{\cal G}_{ik}^{(0)}t_{kl}\Psi_l^{(0)},
\ee
where $\Psi^{(0)}$ is the wave function of the free motion,
\be
\Psi_i^{(0)}(\vep,\veq|\vept,\veqt)=\delta_{i3}\delta(\vep-\vept)\delta(\veq-\veqt),
\ee
in line with the assumed asymptotic.

The solution of Schr{\"o}dinger equation (\ref{schequation}) is
\begin{eqnarray}
&&C(\vept,\veqt)=\frac{f(\vept)
g(\veqt)}{[E-E_X+G_X(E)][E-E_R-T_{ab}(\tilde{p})+G_b(\tilde p,E)]},
\label{Csol}\\[2mm]
&&\chi(\vep|\vept)=\frac{C(\vept,\veqt) f(\vep)+
g(\veqt)\delta(\vep-\vept)}{E-E_R-T_{ab}(p)+G_b(p,E)}, 
\label{chisol}\\[2mm]
&&\vp(\vep,\veq|\vept,\veqt)=\delta(\vep-\vept)
\delta(\veq-\veqt)+\frac{g(\veq)
\chi(\vep|\vept)}{E-T_{ab}(p)-T_{cd}(q)}.
\label{CC}
\end{eqnarray}
Notice that the wave functions are expressed in terms of physical
energies $E_X$ and $E_R$. 

It turns out then that, in practical applications, the quantity of an immediate relevance is the
so-called spectral density, which can be built as
\be
W(E)=\int d^3\tilde p d^3\tilde q \;
|C(\vept,\veqt)|^2\delta\left(E-T_{ab}(\tilde p)-T_{cd}(\tilde q)\right),
\label{WM}
\ee
and which measures the probability to find the bare $|X_0\rangle$ state in the
$|a[cd]\rangle$ continuum --- see Sec.~\ref{motwf}. An explicit calculation gives
\be
W(E)=\frac{1}{2\pi i}
\left[\frac{1}{E-E_X+G_X^*(E)}-\frac{1}{E-E_X+G_X(E)}\right].
\label{Wdef0}
\ee

Comparing Eqs.~(\ref{prr}) and (\ref{Csol}) one can find for the differential rate
\be 
\frac{d{\rm Br}(a[cd])}{dE} \propto \int d^3\tilde p d^3\tilde q\;
|C(\vept,\veqt)|^2\delta\left(E-T_{ab}(\tilde p)-T_{cd}(\tilde q)\right)= W(E).
\label{rate}
\ee

\section{Self-energy calculation}

For energies $E$ small enough one can neglect the form factors --- then
the resulting shapes show universal, model-independent patterns.
In this limit we may write
\begin{eqnarray}
G_b(p,E)=\frac{i}{2}g_{cd}\left[2\mu_q\left(E-\frac{p^2}{2\mu_p}\right)\right]^{l+1/2}
\Theta\left(E-\frac{p^2}{2\mu_p}\right)\nonumber\\+
\frac12(-1)^lg_{cd}\left[-2\mu_q\left(E-\frac{p^2}{2\mu_p}\right)\right]^{l+1/2}
\Theta\left(-E+\frac{p^2}{2\mu_p}\right),\label{Gb1}
\end{eqnarray}
where $l$ is orbital angular momentum in the $cd$ subsystem, and the coupling
$g_{cd}$ is related to the physical resonance width $\varGamma_R$ via
\be
\varGamma_R=g_{cd}(2\mu_qE_R)^{l+1/2}=g_lE_R^{l+1/2},
\label{gldef}
\ee
where, in order to simplify the notations, the shorthand
$g_l=g_{cd}(2\mu_q)^{l+1/2}$ was introduced. Notice that, once we always stay
in the vicinity of the resonance, we retain in Eq.~(\ref{Gb1}) only the part
which changes appreciably in the region of interest, while a nearly constant
(infinite) contribution is omitted in Eq.~(\ref{Gb1}) as it can always be got
rid of through an appropriate redefinition of the physical quantity $E_R$.
We should also stress that
the case of $l \neq 0$ is more subtle, as the form factor $f(q)$ is proportional
to $q^l$, so that the renormalisation procedure requires more than one
subtraction. However, as mentioned here we are primarily interested in the leading
nonanalytical near-threshold contribution from the $[cd]$ self-energy loop,
which is given by expression (\ref{Gb1}).

Thus, the two parameters we treat as input are the energy $E_R$ and the width
$\varGamma_R$. An important dimensionless parameter which governs the dynamics
is
\be
\lambda=\frac{\varGamma_R}{2E_R}.
\ee
We assume $\lambda<1$ --- only then one can speak of the resonance as a constituent\ftnote{3}{For a discussion of
the scattering amplitude for general values of $\lambda$ see
Ref.~\cite{chiralextrapol}.} --- and shall be primarily interested in the case 
$\lambda\ll 1$.

We stick to zero orbital angular momentum in the $ab$ subsystem. It
is then convenient to express $G_X(E)$ in terms of the coupling $g_{ab}$ as
\be
{\rm Re}\left(G_X(E)\right)=\frac12g_{ab}\kappa_{\rm eff}(E),
\quad 
{\rm Im}\left(G_X(E)\right)=\frac12g_{ab}k_{\rm eff}(E),\quad
g_{ab}=8\pi^2\mu_pf^2(0).
\label{kkapdef}
\ee

For a resonance in the $cd$ subsystem one finds
by an explicit evaluation of $G_X(E)$, as given in Eq.~(\ref{GX}), for
the imaginary part
\begin{eqnarray}
k_{\rm eff}(E)&=&\frac{g_l}{2\pi\mu_p}\int_0^{\sqrt{2\mu_pE}}\frac{p^2dp
(E-\frac{p^2}{2\mu_p})^{(2l+1)/2}}{(E_R-E+\frac{p^2}{2\mu_p})^2+\frac{g_l^2}{4}
(E-\frac{p^2}{2\mu_p})^{2l+1}} \ ,
\label{keff}
\end{eqnarray}
where $l$ denotes the angular momentum of the $cd$ system in
its rest frame and $g_l$ was defined in Eq.~(\ref{gldef}). 
The real part is
\be
\kappa_{\rm eff}(E)=\kappa_1(E)+\kappa_2(E)-\kappa_1(E_X)-\kappa_2(E_X),
\label{kapeff}
\ee
where the first term takes the form
\be
\kappa_1(E)=\frac{1}{\pi\mu_p}\int^{\infty}_0p^2dp
\frac{E_R-E+\frac{p^2}{2\mu_p}}{(E_R-E+\frac{p^2}{2\mu_p})^2+
\frac{g_l^2}{4}(E-\frac{p^2}{2\mu})^{2l+1}},
\label{kap1eff}
\ee
and the second term is
\be
\kappa_2(E)=-\frac{g_l}{2\pi\mu_p}\int^{\infty}_{\sqrt{2\mu_pE}}
p^2dp\frac{(\frac{p^2}{2\mu_p}-E)^{(2l+1)/2}}{(E_R-E+\frac{p^2}{2\mu_p})^2+
\frac{g_l^2}{4}\left(E-\frac{p^2}{2\mu}\right)^{2l+1}}.
\label{kappa2full}
\ee
Notice that $\kappa_2(E)$ is suppressed as compared to $\kappa_1(E)$ by a factor
$g_l\propto{\lambda}$, which is small by assumption. We neglect $\kappa_2(E)$ in
what follows throughout the paper. 

For example, for $l=0$ a straightforward evaluation gives
\begin{eqnarray}
k_{\rm eff}(E)
&=&\frac{g_0\sqrt{2\mu_p}}{2}\left(-1+\frac{E_R+
\sqrt{(E-E_R)^2+\frac{g_0^2E}{4}}}{\sqrt{2E_R\left(E_R-E+\frac{g_0^2E}
{8E_R}+\sqrt{(E-E_R)^2+\frac{g_0^2E}{4}}\right)}}\right)
\label{keff0}
\end{eqnarray}
and
\be
\kappa_{\rm eff}(E)\approx\kappa_1(E)-\kappa_1(E_X),
\label{kappa1}
\ee
with
\be
\kappa_1(E)=\mbox{const}+\sqrt{\mu_p}\frac{E-E_R+\frac{g_0^2}{4}
-\sqrt { (E_R-E)^2+
\frac{g_0^2E}{4}}}{\sqrt{E_R-E-\frac{g_0^2}{8}+
\sqrt{(E_R-E)^2+\frac{g_0^2E}{4}}}}.
\ee
Notice that the (infinite) constant is cancelled
in $\kappa_{\rm eff}$, as follows from Eq.~(\ref{kappa1}).

For $l>0$ the integrand in $\kappa_{\rm eff}$ shows, in addition to
the two physical poles, spurious poles in the region $p\gg \sqrt{2\mu_p E}$.  
Their appearance is an artifact of the formalism used [in
particular,  a consequence of neglecting the form factor in
expression (\ref{G2}) for the self-energy of particle $b$]. However, since these
poles are well outside the region of validity of the formalism
discussed, in what follows we neglect their contribution (using the
principal value prescription). Then analytical expressions for the $k_{\rm eff}$ and $\kappa_{\rm eff}$ can be found, in analogy with the case of $l=0$. However, such expressions are bulky, so we do not give them here and resort to numerical calculations instead.

\begin{figure}[t]
\begin{center}
\epsfig{file=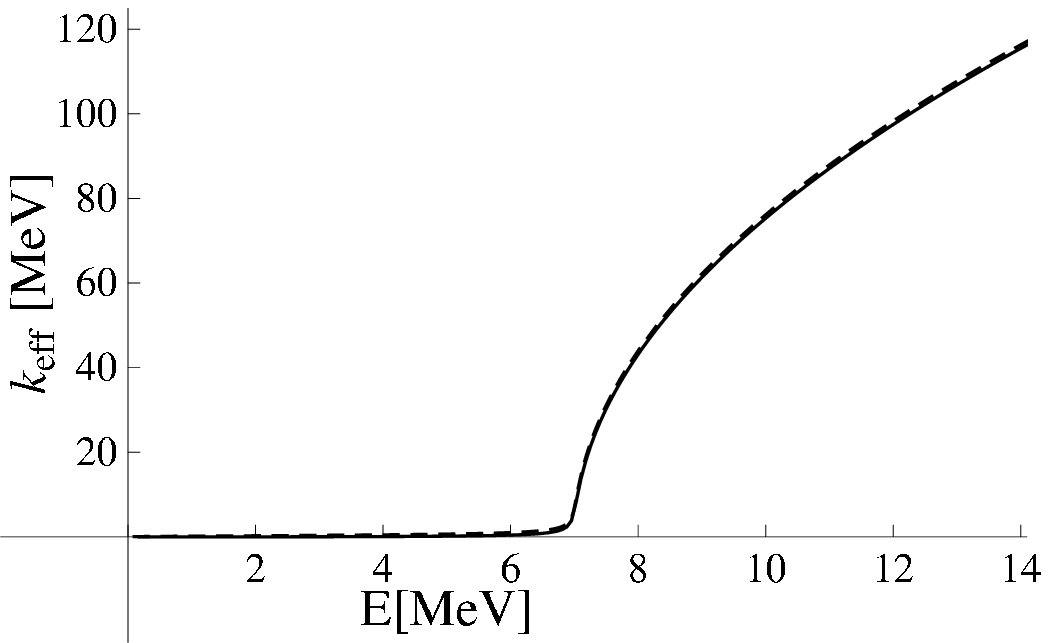,width=7cm}\hspace*{1cm}\epsfig{file=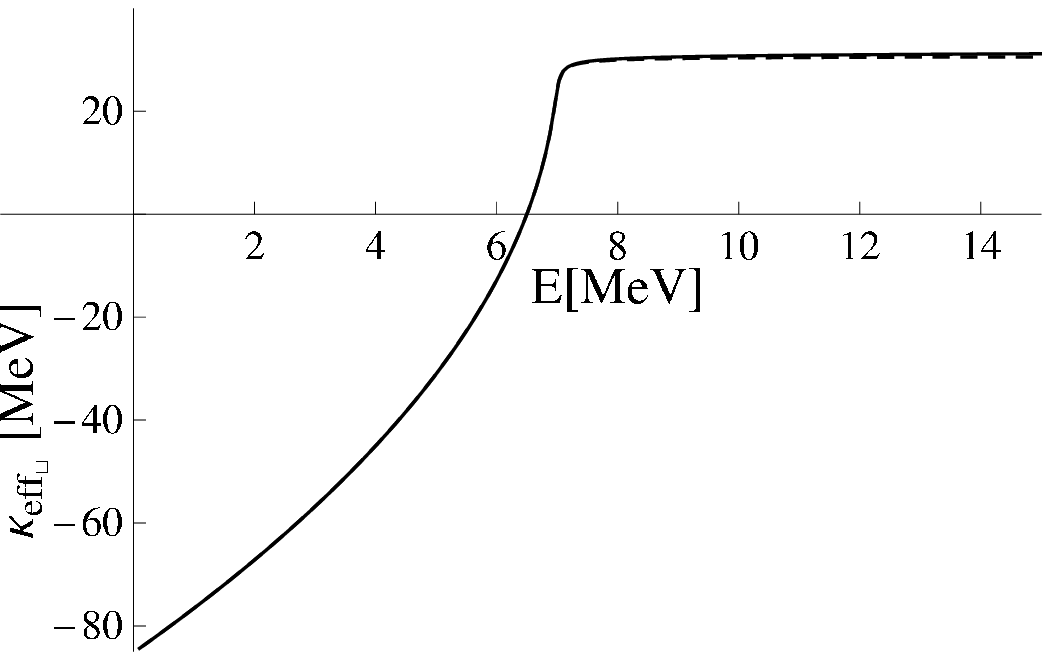,width=7cm}
\caption{The exact results for $k_{\rm eff}(E)$ and $\kappa_{\rm eff}(E)$ for the case $l=0$ versus the
simplified formulae (\ref{kwoolly}) and (\ref{kappawoolly})
with an energy dependent width as proposed in Ref.~\cite{braaten1} for $E_R=7$~MeV and
$\varGamma_R=0.1$~MeV ($\lambda=1/140$). The left plot is for $k_{\rm eff}(E)$, the right plot is
for $\kappa_{\rm eff}(E)$. The $\kappa_2$ term is neglected.}
\label{widthS01}
\end{center}
\end{figure}

\begin{figure}[t]
\begin{center}
\epsfig{file=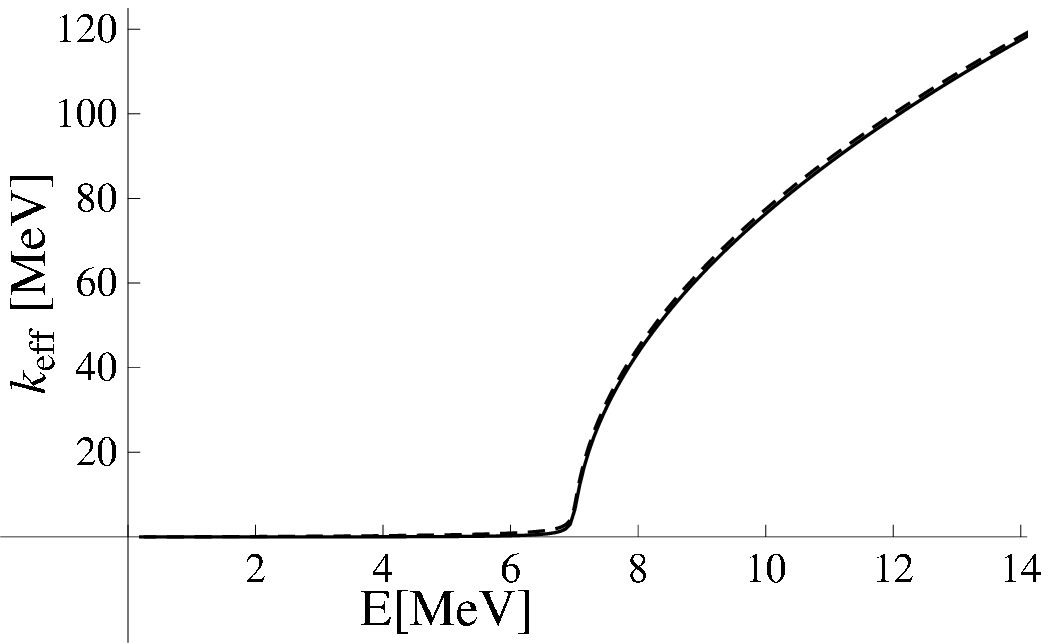,width=7cm}\hspace*{1cm}\epsfig{file=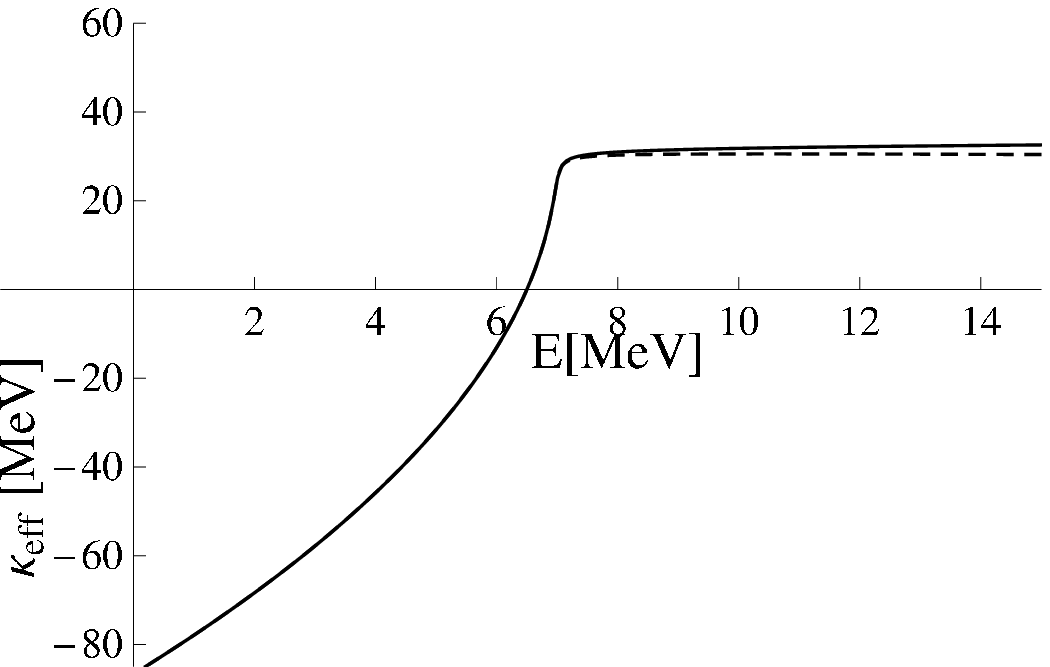,width=7cm}
\caption{The same as in Fig.~\ref{widthS01}, but for $l=1$.}\label{widthP01}
\end{center}
\end{figure}

\begin{figure}[t]
\begin{center}
\epsfig{file=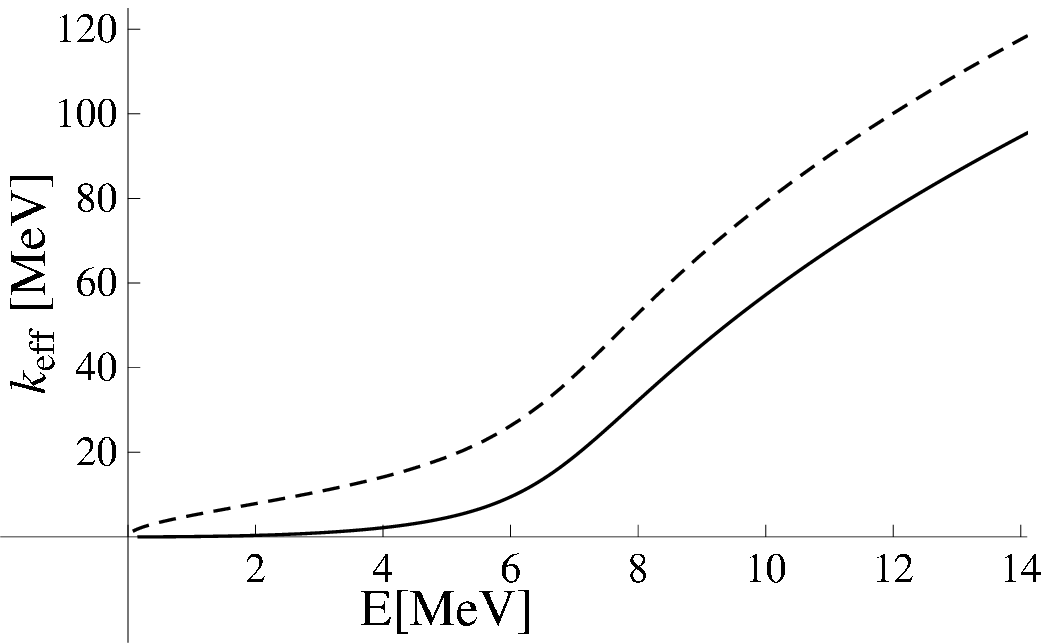,width=7cm}\hspace*{1cm}\epsfig{file=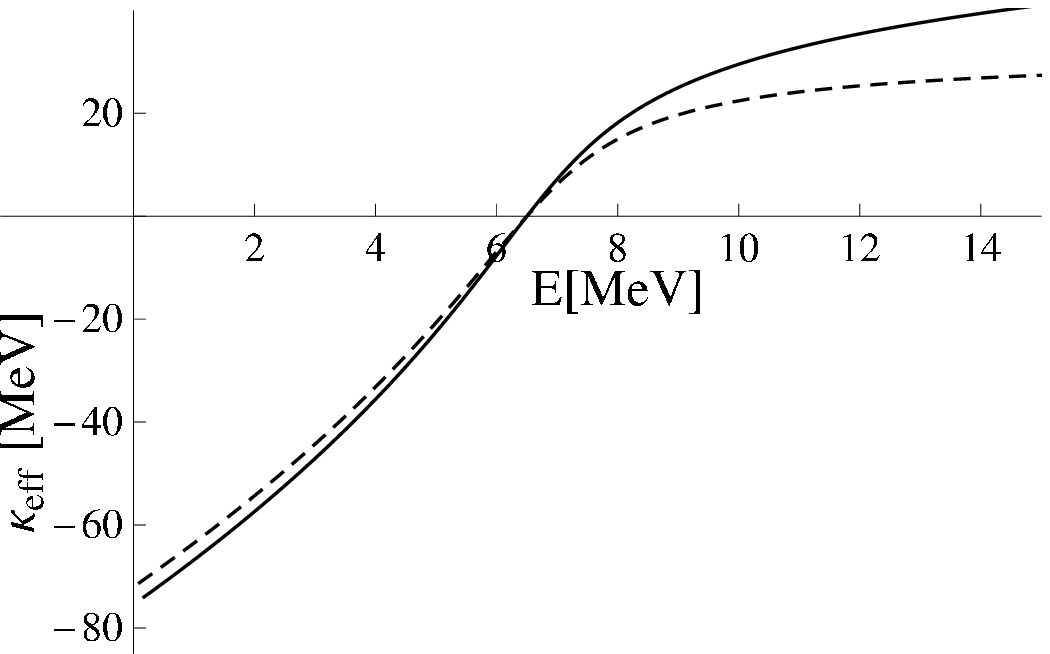,width=7cm}
\caption{The same as in Fig~\ref{widthS01}, but for
$\varGamma_R=3$~MeV ($\lambda=0.2$).}\label{widthS5}
\end{center}
\end{figure}

\begin{figure}[t]
\begin{center}
\epsfig{file=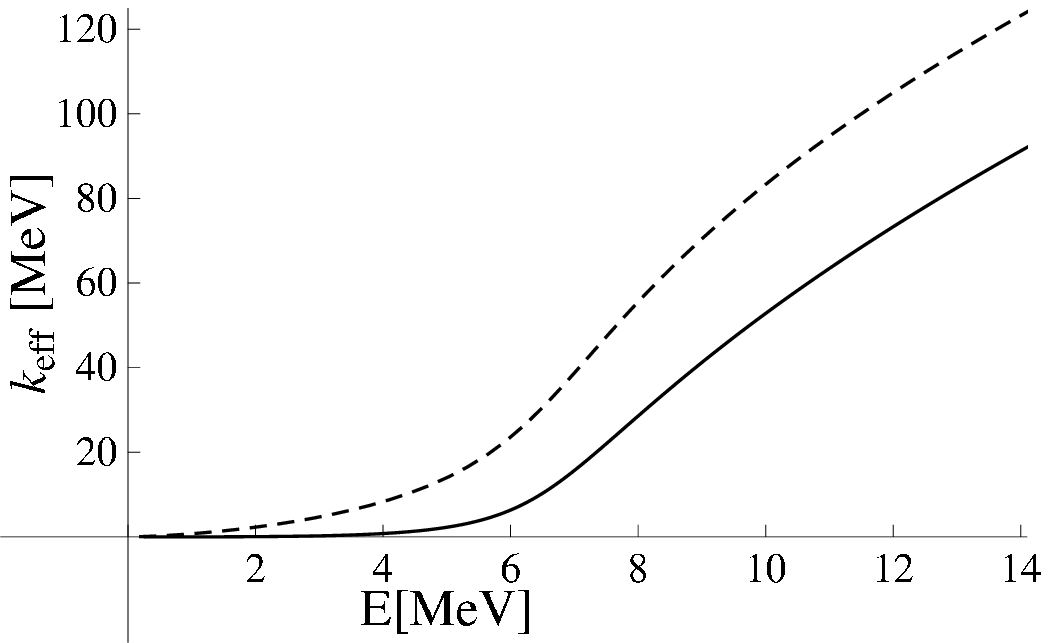,width=7cm}\hspace*{1cm}\epsfig{file=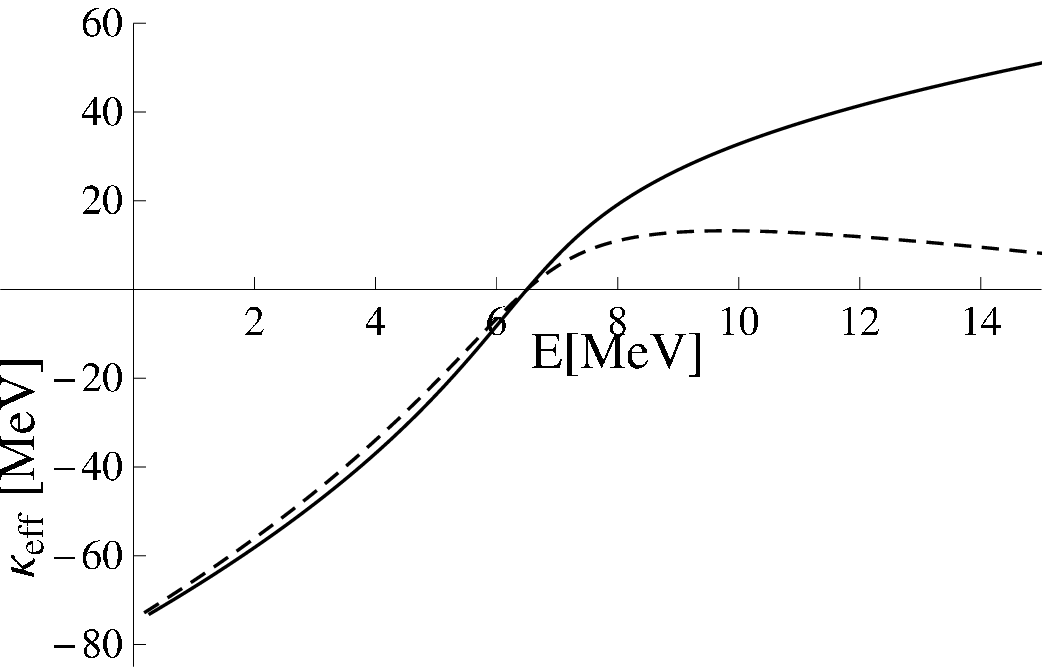,width=7cm}
\caption{The same as in Fig.~\ref{widthS5}, but for $l=1$.}\label{widthP5}
\end{center}
\end{figure}

To demonstrate quantitatively the comparison of Eqs.~(\ref{keff}),
(\ref{kapeff}) versus Eqs.~(\ref{kwoolly}), (\ref{kappawoolly}) --- for
the sake of concreteness in all plots we use the energy-dependent width as proposed in Ref.~\cite{braaten1} --- we choose
two possible parameter sets. As the first example we choose the
parameters as they apply to the case of the $X(3872)$: for the width
of particle $b$ we take an empirical value of order of the $D^*$ state
width, $\varGamma_R=0.1$ MeV. In addition, the mass of the $D^{*0}$
is located at about $E_R=7$ MeV above the $D^0\pi^0$ threshold. This
corresponds to $\lambda=1/140$. The resulting energy dependencies of
both $k_{\rm eff}(E)$ and $\kappa_{\rm eff}(E)$ are given in
Figs.~\ref{widthS01} and \ref{widthP01} for $S$- ($l=0$) and
$P$-waves ($l=1$), respectively. Note that the latter case is relevant
for the $X(3872)$, with the $D^*$ decaying into $D\pi$ in a $P$-wave. 
Since we subtracted both expressions for $\kappa_{\rm eff}(E)$ at the same point $E=E_X$, 
the two curves agree exactly at the subtraction point, which is chosen here as $E_X=E_R-0.5$ MeV.
Clearly, for the parameters relevant for the $X(3872)$ the results
agree perfectly well.  However, already for a width of only about
1~MeV, deviations between the formulae start to become visible. In
Figs.~\ref{widthS5} and ~\ref{widthP5} we compare the two formalisms
for $\varGamma_R=3$~MeV (or $\lambda\simeq 1/5$) with all other
parameters unchanged. Now the difference, especially in $k_{\rm eff}$,
is sizable, mainly as a result of the very different threshold
behaviour. Indeed, at the three-body threshold, Eq.~(\ref{keff}) behaves as
\be
\label{sthres}
k_{\rm eff}(E)\mathop{\approx}
\limits_{E\to 0}\frac12\varGamma_R
\sqrt{\frac{\mu_p}{2E_R}}\left(\frac{E}{2E_R}\right)^{l+2}\propto E^{l+2} \ ,
\ee
in line with the three-body phase space and the centrifugal barrier.

In the meantime, $k^{(0)}_{\rm eff}(E)$, as given by Eq.~(\ref{kwoolly}),
shows a wrong threshold behaviour. If a constant width is used for the
resonance (this case is not shown in the figures), there remains a finite
imaginary part even at the threshold, while for the energy-dependent width (\ref{runwid}) of
Ref.~\cite{braaten1} one finds
\be 
k^{(0)}_{\rm eff}(E)\mathop{\approx}\limits_{E\to 0}
\frac{\varGamma_R\sqrt{2\mu_p E}}{4E_R}\left(\frac{E}{E_R}\right)^l\propto E^{(2l+1)/2}.
\ee 
Thus, instead of the required behaviour according to the three-body phase space, the
approximate result scales as the two-body phase space.

One concludes therefore that, if the resonance is well separated from the three-body threshold,
the simple formulae (\ref{kwoolly}) and (\ref{kappawoolly}) suggested in Ref.~\cite{woolly} may be
used instead of Eqs.~(\ref{keff}) and (\ref{kapeff}). Indeed, it is easy to verify that, in the limit
\be
\lambda\ll 1,\quad \frac{|E_R-E|}{E_R}\ll 1,
\label{constantwidth}
\ee 
our formulae reduce naturally to Eqs.~(\ref{kwoolly}), (\ref{kappawoolly}) [it is an easy task to check this
reduction analytically for the $S$-wave case --- see Eqs.~(\ref{keff0}) and (\ref{kappa1})]. Notice that
in this case the width of particle $b$ can safely be taken as constant in Eqs.~(\ref{kwoolly}) and
(\ref{kappawoolly}) and there is no improvement to use an energy-dependent width 
from Ref.~\cite{braaten1} instead. This is definitely so for the parameters relevant for
the $X(3872)$. However, when investigating particles with broader
constituents, one has to resort to the full formulae (\ref{keff}) and
(\ref{kap1eff}). This applies, in particular, in studies of the
$Y(4260)$. The $Y(4260)$ resides at the
$D_1\bar D$ threshold; there are two $D_1$ states, a narrow one with
a mass of $2420$ MeV and a width of $25$ MeV, and a wide one
with a mass of $2430$ MeV and a width of about $400$ MeV, both
decaying into $D^*\pi$ \cite{PDG}. In the former case one has $\lambda
\approx 0.05$, while in the latter case $\lambda \approx
0.8$.

\section{Modification of the Weinberg formulae}
\label{motwf}

The Weinberg formalism establishes a relation between low-energy
observables, such as the effective-range parameters \cite{weinbergevidence}
or Flatt{\'e} parameters \cite{evidence}, and the nature of a resonance.
In particular, it allows one to quantify the admixture of a genuine compact $X_0$
component and a molecular component in the physical state $X$
in a model-independent way. The formalism was
designed to deal with two-body final states, and in this chapter we address the
question as to how the formulae are modified by the presence of an
unstable constituent.

The central quantity is the spectral density which, in the near-threshold
region, takes the form [see Eqs.~(\ref{Wdef0}) and (\ref{kkapdef})]
\be
W(E)=\frac{1}{2\pi i}
\left[\frac{1}{E-E_X+\frac12g_{ab}\kappa_{\rm eff}-\frac{i}{2}g_{ab}k_{\rm eff}}
-\frac{1}{E-E_X+\frac12g_{ab}\kappa_{\rm eff}+\frac{i}{2}g_{ab}k_{\rm eff}}\right],
\label{w}
\ee
which is nonzero above the lowest relevant threshold (in our case for $E>0$). 
In the absence of three-body bound states it is normalised to unity,
\be
\int^\infty_0W(E)dE=1.
\ee
In the vicinity of the threshold at $E=E_R$ the integral
\be
\int_{E_R-\delta}^{E_R+\delta} W(E)dE
\label{Wdef}
\ee measures the admixture of the bare state for the near-threshold
resonance.  Indeed, if this integral appears of order unity, then one
deals with a genuine quarkonium while a small integral value indicates
a dynamical origin of the resonance.  Therefore, fitting the
production rate (\ref{rate}) with the formula (\ref{w}), one can
estimate the admixture of the bare $X_0$ state in the wave function of
the $X$-resonance in the near-threshold region defined by the
parameter $\delta$ \cite{evidence}. Most naturally, the scale for
$\delta$ is set by the visible width of the resonance, while other
scales, like the distance from the next threshold [$D^+ \bar D^{*-}$
in the case of the $X(3872)$ particle] could also be relevant.

In the narrow-width limit (\ref{constantwidth}) one can proceed
further.  Strictly speaking, with the finite width of particle $b$
taken into account (and in the absence of three-body or two-body bound
states), the physical space is spanned by the genuine three-body
continuum, and the notion of the $ab$ on-shell scattering amplitude
ceases to exist. Nevertheless, for small enough energies and in the
limit (\ref{constantwidth}), the $t$-matrix element $t_{22}$ takes the
form 
\be
t_{22}(E)=\frac{1}{8\pi^2\mu_p}\frac{g_{ab}}{E-E_X+\frac{i}{2}g_{ab}k_{\rm eff}+ \frac12g_{ab}\kappa_{\rm eff}},
\label{t22}
\ee
which can be viewed, for $\varGamma_R \to 0$, as a Flatt{\'e}-type representation for the $t$-matrix in
the $ab$ channel. 

With expression (\ref{t22}) one can study the structure of
singularities in the complex plane. If one of the particles is
unstable, the Riemann surface does not show a two-body but
a three-body cut (if the unstable particle predominantly decays
via a two-particle decay, as is assumed throughout this paper).
As shown above, in the narrow-width limit, one can replace
the expressions for the $k_{\rm eff}(E)$ and $\kappa_{\rm eff}(E)$ by
the simple ones, as given by Eqs.~(\ref{kwoolly}) and
(\ref{kappawoolly}) with a constant width. Then the $t$-matrix poles are 
\be
E_{1,2}=E_X+\frac{1}{2}g_{ab}\kappa_X-\frac14 g_{ab}^2\mu_p\pm
\sqrt{\frac{1}{16}g_{ab}^4\mu_p^2-\frac12 g_{ab}^2\mu_p\left(E_X-E_R+\frac{1}{2}g_{ab}\kappa_X\right)
-\frac{i}{4}g_{ab}^2\mu_p\varGamma_R}, 
\ee 
with $\kappa_X=\kappa^{(0)}_{\rm eff}(E_X)$.
In the limit of a stable
particle $b$, $\varGamma_R=0$, and for $E_X<E_R$, there is a bound state
in the $ab$ system, which corresponds to the upper sign above, 
\be
E_1\to E_B=E_X, 
\ee
and it is straightforward to verify that this pole is situated at the
first sheet of the {\it bona fide} Riemann surface of the genuine
two-body case. For $\varGamma_R>0$ the system couples to the sheet
with $a$, $b$, and $c$ on-shell, and what used to be the physical sheet
for stable constituents, is now the unphysical one. However, the pole $E_1$ 
remains to be the one most relevant for this discussion. In its
 vicinity, $t_{22}$ takes the
form
\be 
t_{22}(E) \to \frac{g_{\rm eff}^2}{E-E_B}, 
\ee 
where 
\begin{eqnarray}
g_{\rm eff}^2&=&\frac{g_{ab}}{8\pi^2\mu_p}\left[1-\frac12g_{ab}
\frac{\partial}{\partial E} \left(\kappa_{\rm eff}+ik_{\rm eff}\right)_
{|E=E_B}\right]^{-1}\nonumber\\
&=&\frac{g_{ab}}{8\pi^2\mu_p}\frac{\sqrt{-2\mu_p(E_B-E_R+\frac{i}{2}\varGamma_R)}}
{\sqrt{-2\mu_p(E_B-E_R+\frac{i}{2}\varGamma_R)}+\frac12\mu_pg_{ab}}.
\label{vertex} 
\end{eqnarray}

In the limit (\ref{constantwidth}) and for $E<E_R$ one has $k_{\rm eff}\to 0$. Then 
\be
W(E) \to
\frac{\delta(E-E_B)}{1-\frac12g_{ab}\frac{\partial}{\partial E}
\sqrt{-2\mu_p(E-E_R)}_{|E=E_B}}={\cal Z}\delta(E-E_B).
\label{z}
\ee
The factor ${\cal Z}$ on the right-hand side of Eq.~(\ref{z}) is nothing but the
probability to find the bare state in the wave function of the bound state.
 
Equation (\ref{z}) allows one to express the coupling $g_{ab}$ in terms of ${\cal Z}$
and $E_B$. Then the vertex $g_{\rm eff}$ takes the famous Weinberg form
\cite{weinbergevidence}
\be
g_{\rm eff}^2 \to \frac{\sqrt{-2\mu_p(E_B-E_R)}}{4\pi^2\mu_p^2}(1-{\cal Z}).
\ee
As it was argued previously~\cite{weinbergevidence}, the given expression is correct
only up to terms of order ${\cal{O}}(\sqrt{2\mu_p (E_R-E_B)}/\beta)$, where $\beta$
denotes the range of forces. 
Our analysis reveals that in case of unstable constituents there are, in
addition to the finite-range corrections, also corrections of order $\lambda$. However, for $\lambda\ll 1$
the Weinberg analysis should still be applicable.

\section{Lineshapes}

We are now in a position to discuss the lineshapes of composite particles 
with unstable constituents. For the plots shown below we
consider the $X(3872)$, as a paradigmatic example, which allows
us to quantify also the effect of the inelastic channels in line
with Refs.~\cite{braaten1,recon}.

We start with the expression for the spectral density, similar to 
the
one given by Eq.~(\ref{w}),
\be
W(E)=\frac{1}{2\pi i}\left[\frac{1}
{E-E_X+G_X^*(E)}-\frac{1}{E-E_X+G_X(E)}\right],
\ee
with
\be
G_X(E)=\frac{1}{2}g_{ab}\kappa_{\rm eff}
+\frac{i}{2}\left(g_{ab}k_{\rm eff}+\varGamma_0\right).
\ee
In this expression we introduced, in addition to what we had before, 
the quantity
$\varGamma_0$ to mimic inelastic channels (for example, the channel 
$\pi\pi J/\psi$ in case of
the $X(3872)$).

Then the differential branching fractions are given by (as mentioned before, we absorb all details of the $X$
production into the constant ${\cal B}$):
\be
\frac{d{\rm Br}(a[cd])}{dE}={\cal B}\frac{1}{2\pi}
\frac{g_{ab}k_{\rm eff}(E)}{(E-E_X+\frac{g_{ab}}{2}\kappa_{\rm 
eff}(E))^2+\frac{1}{4}\left(\varGamma_0+g_{ab}k_{\rm
eff}(E)\right)^2},
\label{elastic}
\ee
for the $a[cd]$ channel ($DD\pi$ channel for the $X(3872)$), and
\be
\frac{d{\rm Br}({\rm inel})}{dE}={\cal B}
\frac{1}{2\pi}
\frac{\varGamma_0}{(E-E_X+\frac{g_{ab}}{2}\kappa_{\rm
eff}(E))^2+\frac{1}{4}\left(\varGamma_0+g_{ab}k_{\rm
eff}(E)\right)^2},
\label{inelastic}
\ee
for the inelastic channel.

\begin{figure}[t]
\epsfig{file=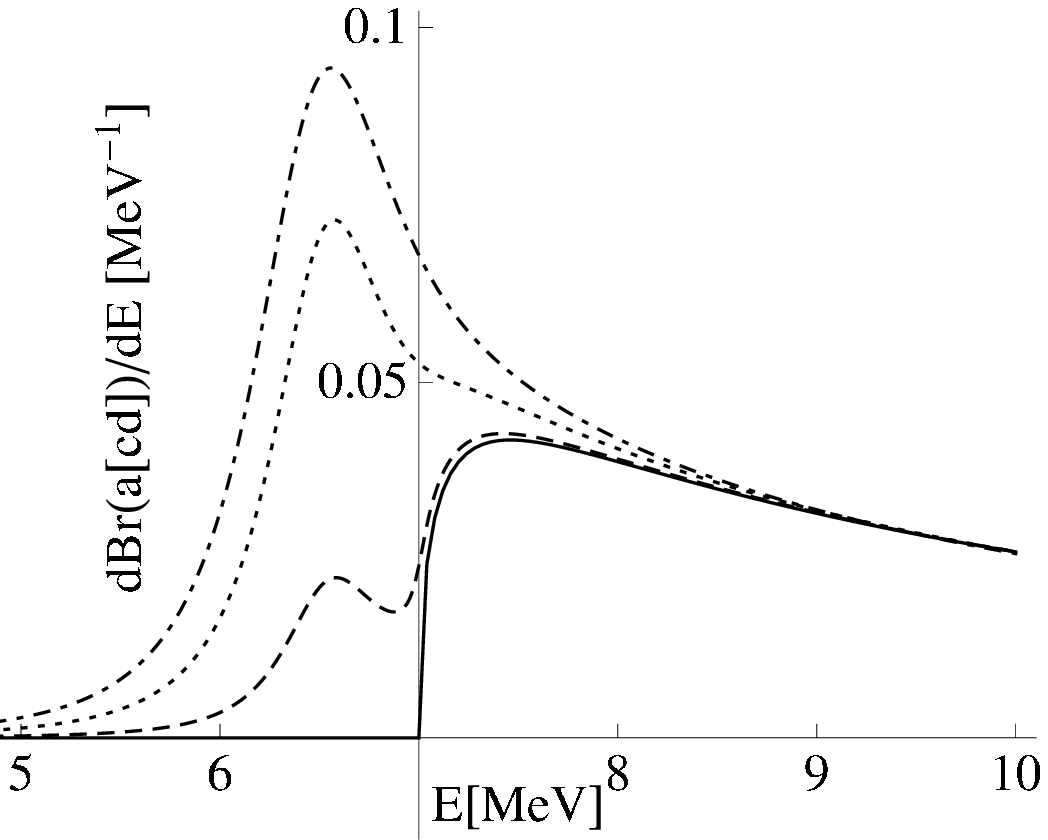,width=7cm}\hspace*{1cm}\epsfig{
file=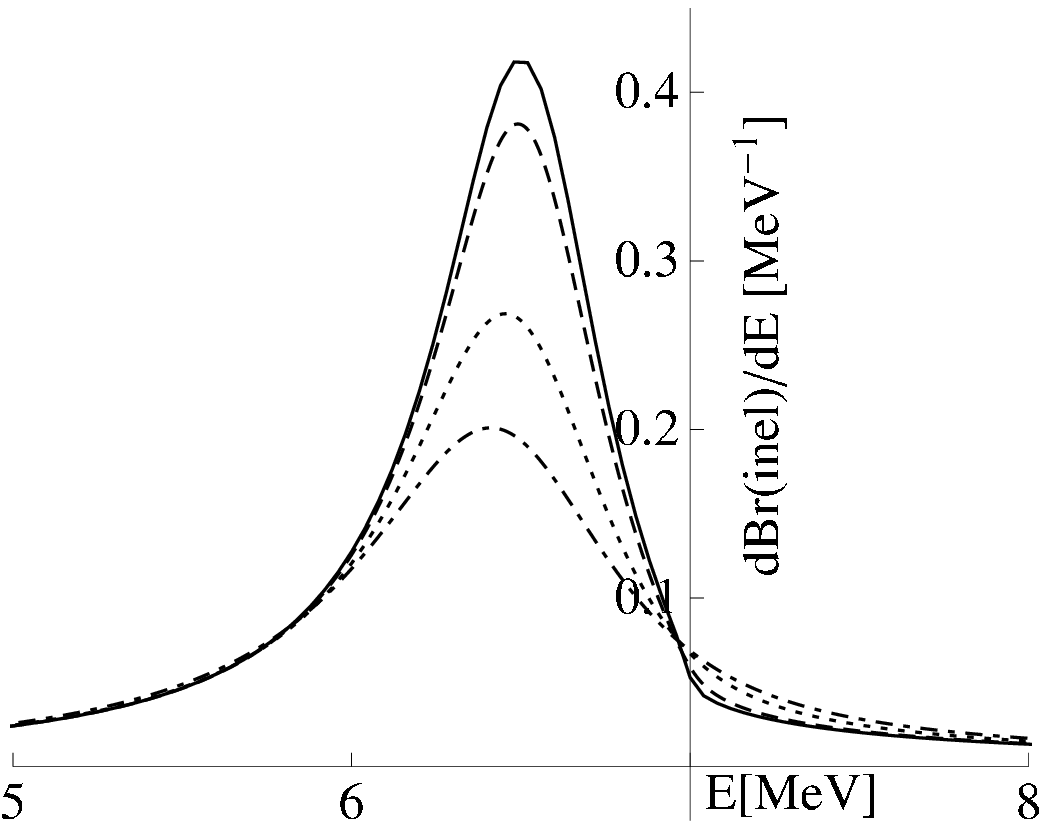,width=7cm}
\caption{The $a[cd]$ production rate [$DD\pi$ rate for the $X(3872)$] (the first plot)
and the inelastic rate [$\pi\pi J/\psi$ for the $X(3872)$] (the second plot) for the
$S$-wave resonance and for the $X$ being a bound state. Here $E_R=7$ MeV and
$\varGamma_R=0$ (solid line),
$\varGamma_R=100$ keV (dashed line), $\varGamma_R=500$ keV (dotted line), and $\varGamma_R=1$ MeV
(dash-dotted line). Notice that, since we are interested in the lineshapes, the above rates are plotted for 
${\cal B}=1$.}\label{BrS1}
\epsfig{file=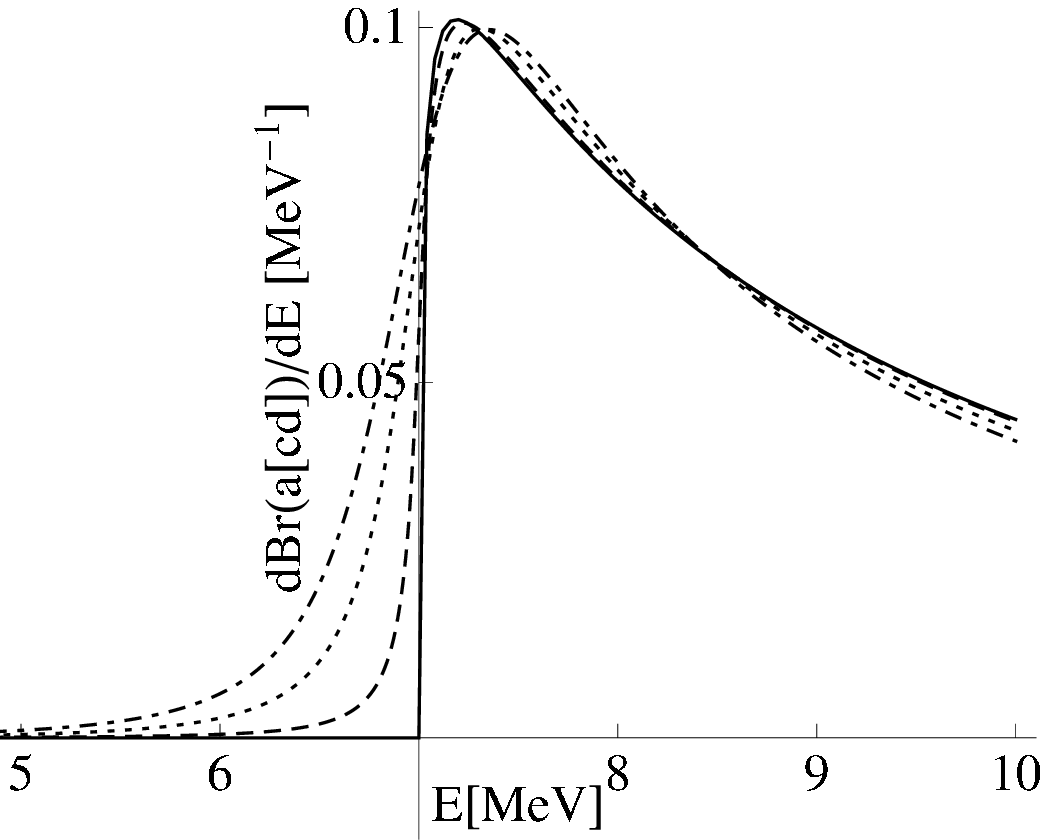,width=7cm}\hspace*{1cm}\epsfig{
file=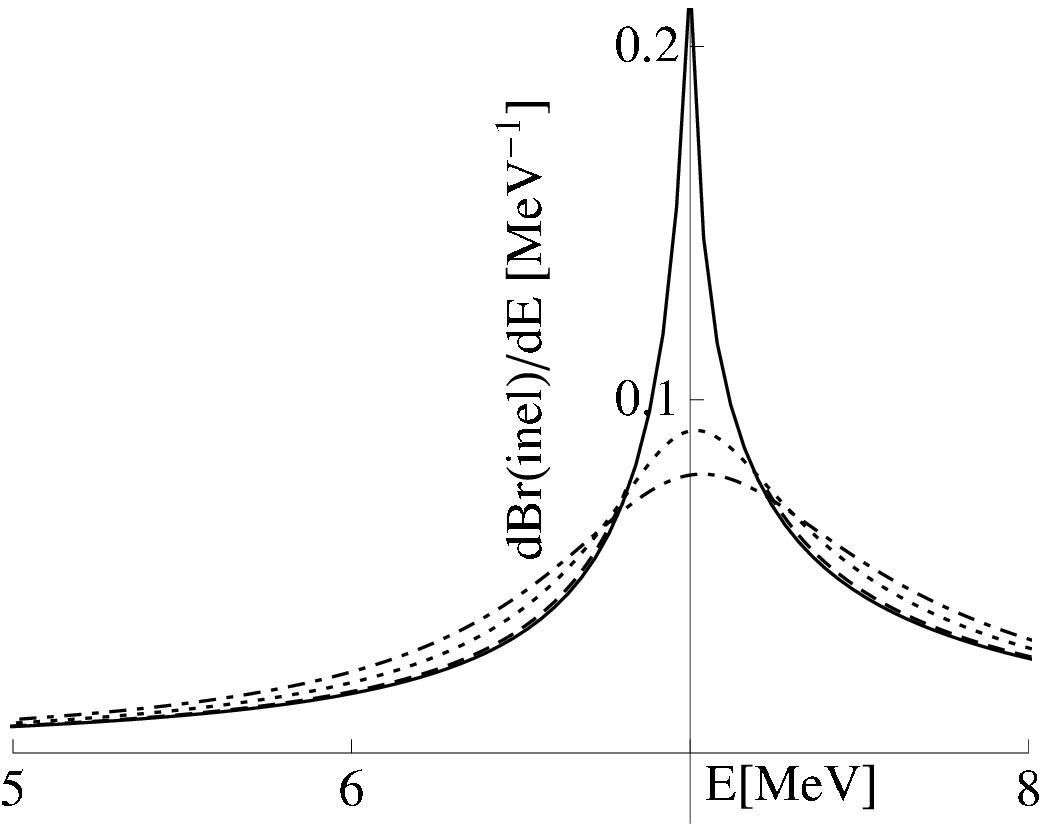,width=7cm}
\caption{The same as in Fig.~\ref{BrS1}, but for the $X$ being a virtual state.}\label{BrS2}
\end{figure}

\begin{figure}[t]
\epsfig{file=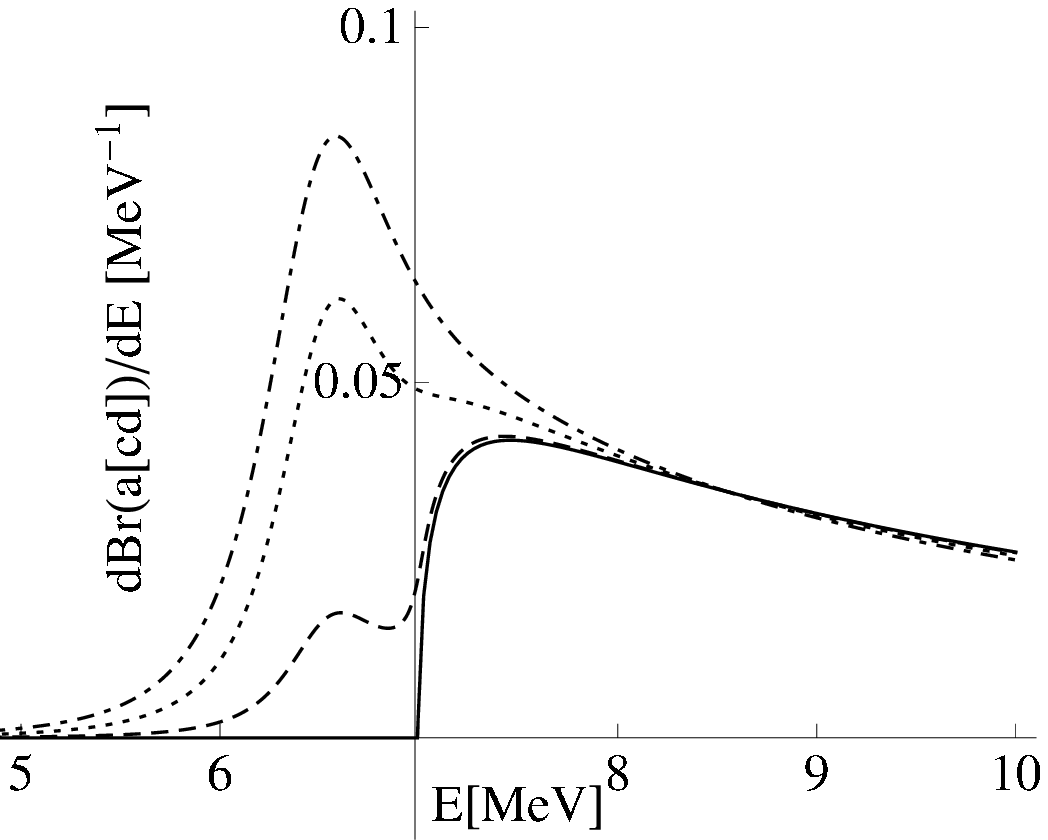,width=7cm}\hspace*{1cm}\epsfig{
file=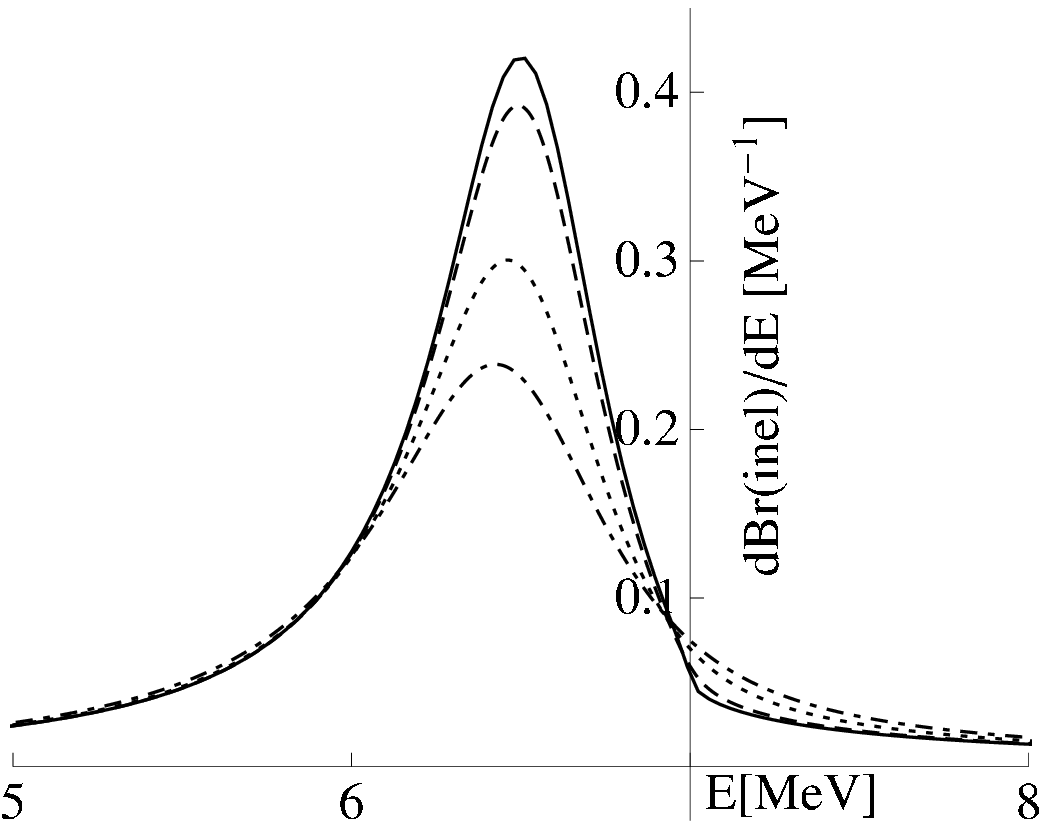,width=7cm}
\caption{The same as in Fig.~\ref{BrS1}, but for the $P$-wave resonance.}\label{BrP1}
\epsfig{file=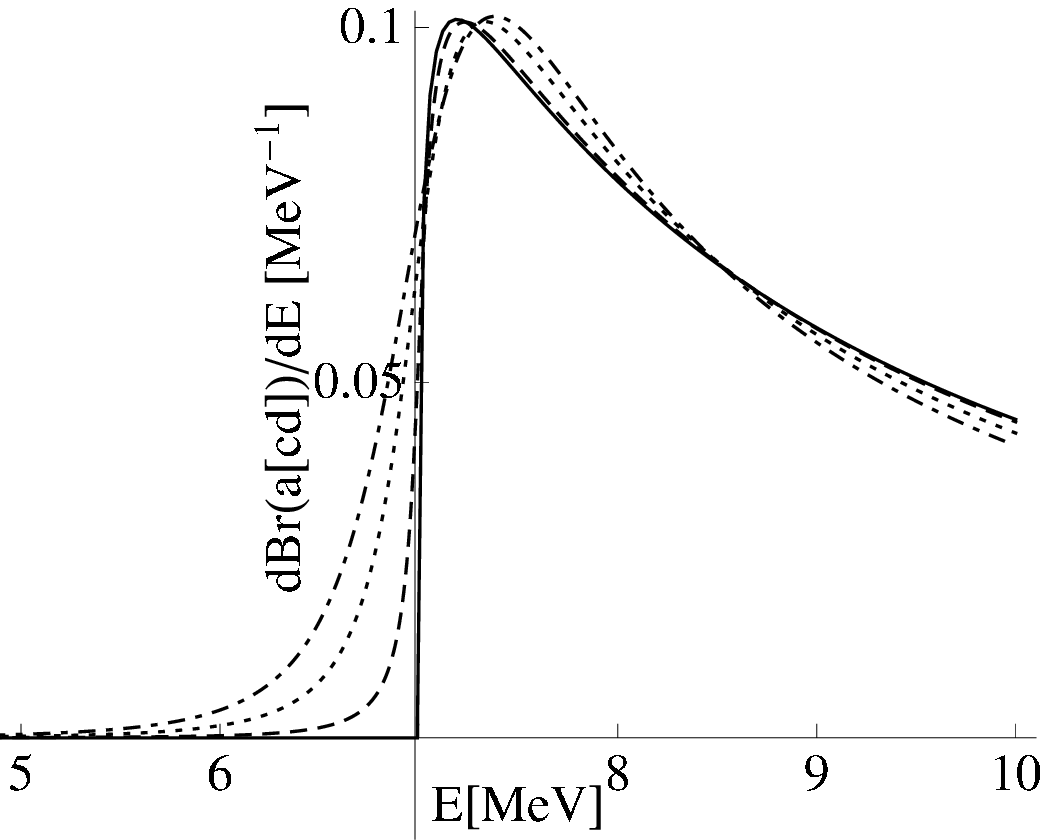,width=7cm}\hspace*{1cm}\epsfig{
file=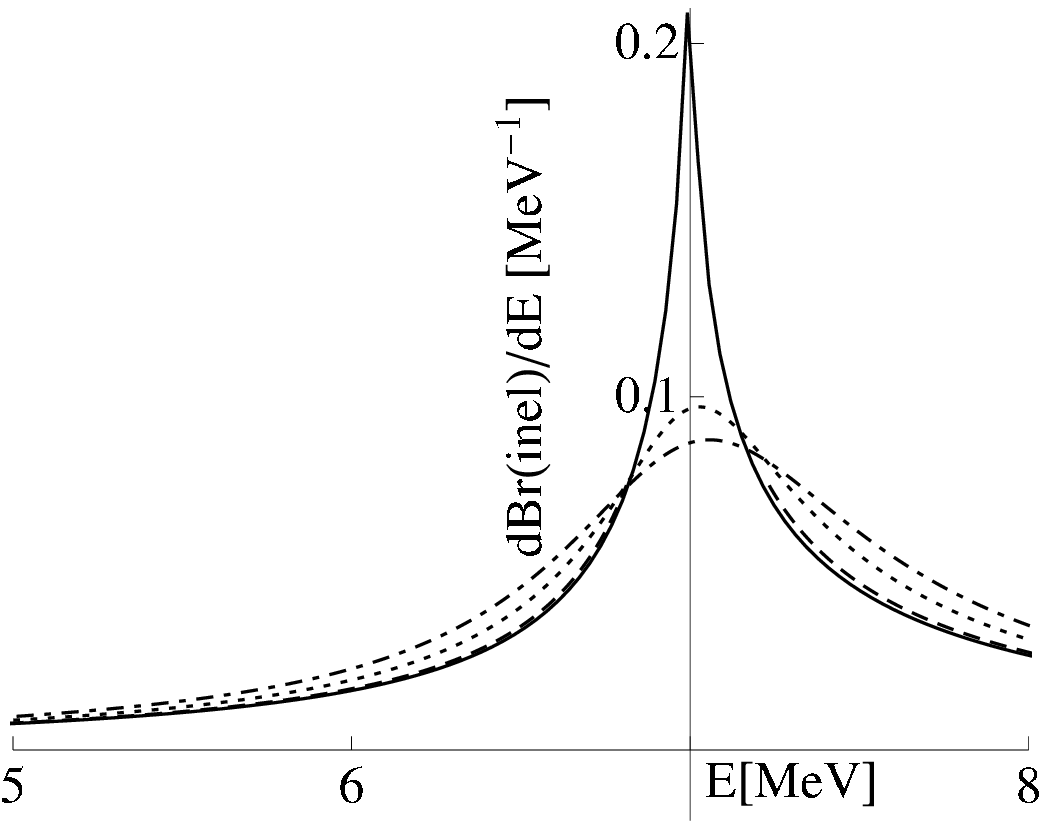,width=7cm}
\caption{The same as in Fig.~\ref{BrS2}, but for the $P$-wave resonance.}\label{BrP2}
\end{figure}

To illustrate the effect of the finite width of one of the constituents we investigate the above-mentioned
differential branching ratios for both the $S$- and $P$-wave case and for
both bound and virtual $X$ particles. However, before we proceed,
let us remind the reader of the definition of  bound and
virtual states and show their relation to the parameters introduced above.
Consider the case when all inelasticities are switched off. Then $k_{\rm eff}(E)$ and 
$\kappa_{\rm eff}(E)$ are given by the simple formulae (\ref{ksim}) and
(\ref{effmomgam0}). To distinguish between bound- and virtual-state 
situations it is convenient to consider the poles of 
the denominator $E-E_X+G_X(E)$ in the $k$-plane, where 
$k=\sqrt{2\mu(E-E_R)}$. For $E_X<E_R$ these poles are   
\be
k_1=i\sqrt{2\mu_p(E_R-E_X)},\quad k_2=-i\sqrt{2\mu_p(E_R-E_X)}-ig_{ab}\mu_p.
\ee
The first pole is situated on the first sheet of the genuine two-body 
Riemann surface, and corresponds to the 
bound state with the energy 
$E_X$, and the second pole is at the unphysical sheet. The scattering 
length in the $ab$ channel is positive for this case. On the contrary, 
for 
$E_X>E_R$ the poles are:
\be
k_{1,2}=-i\frac{g_{ab}\mu_p}{2} \pm i\sqrt{\frac{g^2_{ab}\mu^2_p}{4}
-2\mu_p(E_X-E_R)}.
\ee
Both poles are located on the unphysical sheet, and the scattering length is 
negative. 
For $g_{ab}^2\mu_p>8(E_X-E_R)$ the poles are 
on the imaginary axis, and we deal with a virtual state. 
For small inelasticities, the scattering length acquires a small imaginary
part, so that we distinguish between the bound and the virtual states by the sign of the real part of $a$. 
Thus we plot in Figs.~\ref{BrS1}-\ref{BrP2} the differential branching ratios 
(\ref{elastic}) and (\ref{inelastic}) for both the $S$- and $P$-wave case and $E_X<E_R$ (bound state) and $E_X>E_R$
(virtual state). In particular, for the bound-state case we take
\be
E_R=7~\mbox{MeV},\quad g_{ab}=0.1, \quad E_X=E_R-0.5~\mbox{MeV},\quad \varGamma_0=1.5~\mbox{MeV}.
\label{set}
\ee 
This set of parameters is similar to the one obtained in Ref.~\cite{xnew} from 
the fitting Belle data on the $D^0\bar{D}^0\pi^0$ and $ \pi^+\pi^-J/\psi$ decay 
modes of the $X(3872)$ particle. 

As one can see from Figs.~\ref{BrS1} and \ref{BrP1}, for an increasing
width of the constituent, the lineshape in the $a[cd]$ channel starts
to develop a peak below the nominal threshold, since the finite width
of the unstable constituent allows for an overlap with the
subthreshold pole. If the width is not large, the lineshape above
threshold is not affected while, with an increase of the width of
particle $b$, the structure present below the nominal $ab$ threshold
starts to grow very quickly, and it becomes broader. Indeed, already
for a width of 1~MeV, the structure below the nominal threshold
has grown in magnitude and width so much that the whole lineshape
is just a single broad bump, however, with a noticeable asymmetry
letting it deviate significantly from the standard Breit--Wigner form. The
structure for the inelastic channel decreases rapidly with the
increase of the width of particle $b$. The mentioned kind of asymmetry,
found experimentally in the spectral distribution for the $Y(4660)$
observed in $\psi'\pi\pi$~\cite{Y4660exp}, was interpreted in
Ref.~\cite{ourY4660}, fully in line with the 
reasoning presented in this work, as a clear signal for a dominant molecular
$f_0(980)\psi'$ component in the $Y(4660)$. In addition, the
same kind of asymmetry was predicted
to  show up from an analogous $\eta_c'f_0(980)$ bound system
in the $\eta_c'\pi\pi$ invariant mass distribution,
yet to be found~\cite{ourY2}.

For the virtual-state case we take the same values for the $g_{ab}$
and $\varGamma_0$ as for the bound-state case --- see Eq.~(\ref{set}) --- but
put $E_X=E_R+0.5$~MeV.  Then, as seen from Figs.~\ref{BrS2} and
\ref{BrP2}, the effect of the finite width on the $a[cd]$ lineshape is
less dramatic since, if the $X$ is a virtual state, no pole develops
below the threshold. This is already an unambiguous decisive difference
between the bound state and the virtual state. The inelastic rate
decreases with the increase of the width of particle $b$, and the cusplike
lineshape, present in the zero-width limit, smoothens, so that the shape
for the virtual state becomes hardly distinguishable from that for the
bound state, however, with a shifted peak.
 
It follows from the reasoning of Ref.~\cite{evidence} that
the height of the hump above the nominal $ab$ threshold contains
important information on the nature of the state. We would
therefore like to repeat here the argument given in 
Refs.~\cite{braaten2,xnew} that it is dangerous to apply a kinematic
fit to the data, which moves all relevant signals to above threshold: in this
way the signal above the nominal threshold would
be risen at the cost of the signal below the threshold. Therefore, this procedure
might produce data that call for a very different interpretation of the $X$
state than that for the original data set.

\section{Interference}

In this chapter we discuss one more effect which plays a role for the
lineshapes, namely the interferences in the final states
\cite{voloshinint}. 
For example, the $X(3872)$ decays in equal parts via 
$\bar{D}^0D^{0*}\to\bar{D}^0[D^0\pi^0]$ or 
$D^0\bar{D}^{0*}\to D^0[\bar{D}^0\pi^0]$ and,
clearly, the final states of both decay chains interfere.

If taken naively, the interference can be taken into account through the modification [see Eq.~(\ref{Csol})]
\be
C(\tilde{\bm p},\tilde{\bm q})\to\tilde 
C(\tilde{\bm p},\tilde{\bm q})=\frac{1}{\sqrt{2}}(C(\tilde{\bm p},\tilde{\bm q})
+C({\bar{\bm p}},{\bar{\bm q}}))
\ee
in Eq.~(\ref{WM}), where $\bar{\bm p}$ and $\bar{\bm q}$ are known linear
combinations of the momenta $\tilde{\bm p}$ and $\tilde{\bm q}$:
\be
\bar{\bm p}=-\frac{m_a}{m_c+m_d}\tilde{\bm p}+\tilde{\bm q},\quad
\bar{\bm q}=\frac{m_a}{m_c+m_d}\tilde{\bm q}+\frac{(m_a+m_c+m_d)m_d}{(m_c+m_d)^2}
\tilde{\bm p}.
\ee

The phase-space integration measure is invariant under the variable change
$\{\tilde{\bm p},\tilde{\bm q}\}\to\{{\bar{\bm p}},{\bar{\bm q}}\}$ and the modified 
Eq.~(\ref{WM}) reads
\be
W(M)=\int d^3\tilde p d^3\tilde q\left[|C(\tilde{\bm p},\tilde{\bm  q})|^2+
{\rm Re}(C(\tilde{\bm p},\tilde{\bm  q})C^*({\bar{\bm p}},{\bar{\bm q}}))\right]
\delta\left(M-m_a-m_c-m_d-T_3(\tilde p,\tilde q)\right),
\label{WM2}
\ee
where the second term in the square brackets describes the interference.
It is clear then that the interference simply produces an extra contribution to be added to $k_{\rm eff}(E)$, which 
we denote as $k_{\rm int}(E)$ and which can be extracted from Eq.~(\ref{WM2}) in the form
\be
\label{kintdef}
k_{\rm int}(E)=k_{\rm eff}(E)\frac{\ds\int d^3\tilde p d^3\tilde q\;{\rm Re}(C(\tilde{\bm p},
\tilde{\bm  q})C^*(\bar{\bm p},\bar{\bm q}))
\delta\left(M-m_a-m_c-m_d-T_3(\tilde p,\tilde q)\right)}{\ds\int d^3\tilde p d^3\tilde q|C(\tilde{\bm p},\tilde{\bm 
q})|^2
\delta\left(M-m_a-m_c-m_d-T_3(\tilde p,\tilde q)\right)}.
\ee
Therefore, in leading order in the interference, the $a[cd]$ rate can be written in the form of 
Eq.~(\ref{elastic}) but with the $k_{\rm eff}(E)$ in the numerator substituted by the sum 
$k_{\rm eff}(E)+k_{\rm int}(E)$:
\be
\frac{d{\rm Br}(a[cd])}{dE}={\cal B}
\frac{1}{2\pi}\frac{g_{ab}(k_{\rm eff}(E)+k_{\rm int}(E))}{(E-E_X+\frac{g_{ab}}{2}\kappa_{\rm
eff}(E))^2+\frac{1}{4}\left(\varGamma_0+g_{ab}k_{\rm
eff}(E)\right)^2}.
\label{DDpi}
\ee 

Clearly, taking the interference into account in the
numerator only while neglecting it in the denominator (higher orders in the interference)
leads to a violation of unitarity, which can be cured by solving the full
coupled-channel problem with the effects of dynamical building blocks
of the constituents taken into account from the very beginning. However, bearing in mind a very moderate effect of the interference over the lineshapes, we resort to a much simpler approach which amounts to taking $k_{\rm int}(E)$ into account both in the numerator and in the denominator simultaneously:
\be
\frac{d{\rm Br}(a[cd])}{dE}={\cal B}\frac{1}{2\pi}
\frac{g_{ab}(k_{\rm eff}(E)+k_{\rm int}(E))}{(E-E_X+\frac{g_{ab}}{2}
\kappa_{\rm eff}(E))^2+\frac{1}{4}\left(\varGamma_0+g_{ab}(k_{\rm eff}(E)+k_{\rm int}(E))\right)^2}.
\label{DDpiimproved}
\ee
We call this an improved interference. 

\begin{figure}[t]
\epsfig{file=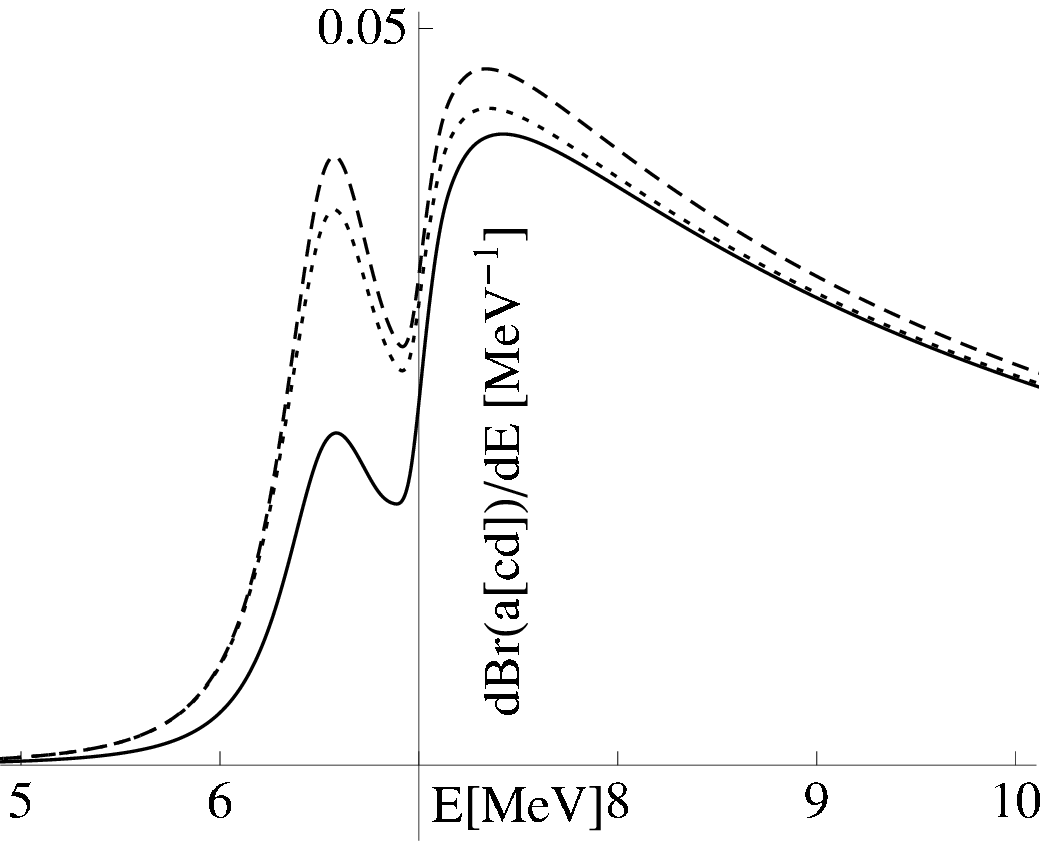,width=7cm}\hspace*{1cm}\epsfig{
file=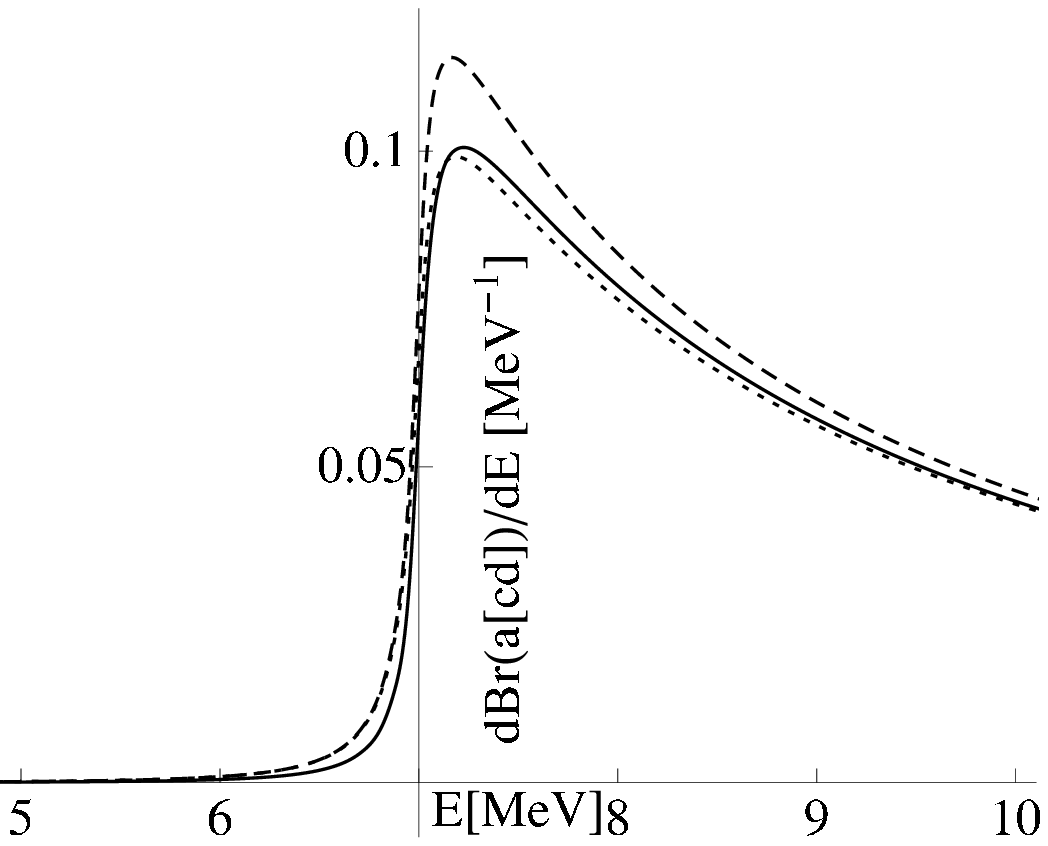,width=7cm}
\caption{The $a[cd]$ state production rate for the $S$-wave resonance for bound-state case (the
first plot) and for the virtual-state case (the second plot). Solid line: no interference; dashed
line: naive interference; and dotted line: improved interference. As before, we set 
${\cal B}=1$.}\label{BrS1v2}
\end{figure}

In Fig.~\ref{BrS1v2} we plot the rate $d{\rm Br}(a[cd])/dE$ for
an $S$-wave resonance and for the following three cases: (i) no
interference [as given by Eq.~(\ref{elastic})], (ii) naive
interference [as given by Eq.~(\ref{DDpi})], and (iii) improved
interference [as given by Eq.~(\ref{DDpiimproved})]. 
For definiteness, we fix all parameters compatible with the $X(3872)$ charmonium, so that the parameters
used are
\be 
E_R=7~\mbox{MeV},\quad g_{ab}=0.1, \quad E_X=E_R \mp
0.5~\mbox{MeV},\quad \varGamma_0=1.5~\mbox{MeV},\quad
\varGamma_R=100~\mbox{keV}, 
\ee 
where upper (lower) sign in $E_X$ corresponds to bound(virtual)-state case. Finally, we set $m_a=m_c=m_0$ and
$m_d=m_\pi$, with $m_0$ and $m_{\pi}$ being the masses of the $D^0$ and $\pi^0$ mesons, respectively. 

The corresponding plots for the $P$-wave exhibit exactly the same 
behaviour as for the $S$-wave and are omitted here. 

As one could anticipate,
the net effect of the interference is an increase of the rate in the
region of the bound-state peak, while the above-threshold rate remains
practically the same.

A proper account for the interference obviously requires a more
rigorous treatment of the coupled-channel problem. Indeed, in the case
of the $X(3872)$, a potentially important piece of interaction is
omitted in the Hamiltonian (\ref{H}), which is responsible for the
$D^0\bar{D}^{0*}$--$\bar{D}^0D^{0*}$ transition, namely the  pion
exchange. This interaction is often considered as the main binding
mechanism in the $D^0\bar{D}^{0*}$ system
\cite{voloshin,tornqvist,swanson}, and the $X$ is generated
dynamically, similarly to the deuteron. It was noticed, however, that,
as the nominal threshold for $D^0\bar{D}^{0*}$ is only about 7 MeV
larger than the three-body $D^0\bar{D}^0\pi^0$ threshold, the pion may
go on-shell and, as a result, such a binding may not be strong enough
(see, for example, Ref.~\cite{suzuki}).  It was argued in
Ref.~\cite{pions} that the effect of the $\pi^0$ exchange for the
$D^0\bar{D}^{0*}$ system can be treated perturbatively, in contrast to
the two-nucleon system. If it is indeed the case, the expression
(\ref{DDpi}) for the $D^0\bar{D}^0\pi^0$ rate should be appropriate.
This statement is further supported by the observation that the rates
described with the ``naive'' and the ``improved'' prescriptions for
the interference do not differ a lot.

\section{Summary}

In this paper we investigated the effect of the finite width of a
constituent of a composite state on its lineshape. In a
coupled-channel formalism we derived exact formulae for such
lineshapes and compared them to simple prescriptions found in the
literature. We found that such simple prescriptions work only
for extremely narrow constituents located far away from the three-body
threshold characterised by  $\lambda={\varGamma_R}/({2E_R})\ll 1$,
where $\varGamma_R$ denotes the width of the constituent and
$E_R$ its resonance energy, measured with respect to the three-body
threshold. In this regime the resonance width can safely be
treated as energy independent. The
famous charmonium state $X(3872)$ gives an example of such a safe
situation with $\lambda\ll 1$ in view of an extremely small width of the $D^*$
meson. However, for an increasing width of the constituent (and
correspondingly an increasing value of $\lambda$), the exact
formulae start to deviate significantly from the simplified ansatz. The
situation cannot be improved by using an energy-dependent width
for the constituent instead
of a constant width --- the three-body threshold behaviour gets
spoilt in this case, while the near-two-body threshold behaviour is
not improved. We argue therefore that the exact formulae have to be
employed in such a situation.

The effect of the finite width of a constituent on the lineshape
in the elastic channel of the
composite state $X$ is quite moderate, if the $X$ is a virtual
state. However, for the $X$ being a bound state, a nonnegligible bound-state
peak starts to develop below the nominal threshold for an increasing
constituent width, as already stressed in
Ref.~\cite{braaten1}. For sufficiently large values of the width, the
bound-state peak becomes broad enough to distort the lineshape both below and
above the threshold, so that the resulting lineshape takes the form of a broad
hump that one can hardly distinguish in shape from the
virtual-state situations. However, the peak position appears to be shifted,
since for the virtual state the distribution is peaked above the nominal
two-body threshold while for the bound state it is peaked below. Clearly,
to distinguish the two experimentally calls for a quite high resolution.

The inelastic production
rate is less affected by the finite-width effects. However, only for extremely
small values of the width one is able to distinguish between the threshold
cusp scenario, in case of the virtual state, and a smooth below-threshold
peak, in case of the bound state. For an increasing width, a smeared broad
structure is developed in both cases. This makes it difficult to
distinguish between the two scenarios, although the peaks of the two
distributions appear to be shifted also in this case.  It is important
to stress that the distributions turn out to
be very asymmetric --- inconsistent with a Breit--Wigner shape --- in both
cases. This asymmetry can be taken as a strong signal towards a composite
two-hadron structure in distinction to a genuine, compact quark state.

As a related problem, we studied the interference in the final
state. For the sake of definiteness, we focused on the case of the
$X(3872)$ charmonium and fixed all free parameters accordingly. The
conclusion we find is that, while the effect of the interference on
the elastic lineshape is indeed sizable for the bound state and below
the nominal threshold, in line with the findings reported in the
literature, this effect is negligible for the bound state above the
threshold as well as for the virtual state both below and above the
threshold. However, for an increasing width one is to include the
effects of interference. Furthermore, the
naive inclusion of the interference meets severe problems with 
unitarity, so that, while for intermediate values of the width, an
improved interference suggested in this paper should be used, for
a width large enough one is finally forced to solve the full
coupled-channel problem with the effects of dynamical building blocks
of the constituents taken into
account from the very beginning.

We treat our findings as an important step in building a universal
lineshape parameterisation scheme suitable for fitting experimental
data for various near-threshold resonances and thus we believe that
our insights will prove useful for the analysis of future experiments.

\begin{acknowledgments}
We would like to thank Eric Braaten, Feng--Kun Guo, and Bruce Yabsley for valuable comments.
This work was supported in parts by funds from the Helmholtz
Association (grants VH-NG-222, VH-VI-231), by the DFG (grants SFB/TR 16
and 436 RUS 113/991/0-1), by the EU HadronPhysics2 project, by the RFFI (grants RFFI-09-02-91342-NNIOa and
RFFI-09-02-00629a), and by the Presidential programme for support of the leading scientific schools (grant
NSh-4961.2008.2).
Work of Yu. K. and A. N. was supported by the State Corporation of Russian
Federation ``Rosatom.'' A. N. would also like to acknowledge the
support of the grants PTDC/FIS/70843/2006-Fi\-si\-ca, and of
the nonprofit Dynasty foundation and ICFPM.
\end{acknowledgments}

\end{document}